\documentclass[proof]{WileyASNA-v2}

\usepackage{amssymb}
\usepackage{graphicx}
\usepackage{times}
\usepackage{amsmath,siunitx}
\usepackage{multicol,longtable}
\usepackage{supertabular}
\usepackage{hyperref}
\usepackage{threeparttable}
\usepackage{stix}

\articletype{ORIGINAL ARTICLE}%

\received{09 November 2020}
\revised{--}
\accepted{07 December 2020}

\begin{document}

\title{A search for runaway stars in twelve Galactic supernova remnants\protect\thanks{Based on service mode observations obtained from ESO-VLT in project 0100.D$-$0314 and NAOJ-Subaru in project S18B0195S.}}

\author[1]{Oliver Lux*}

\author[1]{Ralph Neuh\"auser}

\author[1]{Markus Mugrauer}

\author[1]{Richard Bischoff}

\authormark{LUX \textsc{et al.}}

\address[]{\orgdiv{Astrophysical Institute and University Observatory}, \orgname{Friedrich Schiller University Jena}, \orgaddress{\state{Thuringia}, \country{Germany}}}

\corres{*Oliver Lux, Schillerg\"a{\ss}chen 2, 07745 Jena. \email{oliver.lux@uni-jena.de}}

\abstract{Runaway stars can result from core-collapse supernovae in multiple stellar systems. If the supernova disrupts the system, the companion gets ejected with its former orbital velocity. A clear identification of a runaway star can yield the time and place of the explosion as well as orbital parameters of the pre-supernova binary system. Previous searches have mostly considered O- and B-type stars as runaway stars because they are always young in absolute terms (not much older than the lifetime of the progenitor) and can be detected up to larger distances. We present here a search for runaway stars of all spectral types. For late-type stars, a young age can be inferred from the lithium test. We used \textit{Gaia} data to identify and characterise runaway star candidates in nearby supernova remnants, obtained spectra of 39 stars with UVES at the VLT and HDS at the Subaru telescope and found a significant amount of lithium in the spectra of six dwarf stars. We present the spectral analysis, including measurements of radial velocities, atmospheric parameters and lithium abundances. Then we estimate the ages of our targets from the Hertzsprung-Russell diagram and with the lithium test, present a selection of promising runaway star candidates and draw constraints on the number of ejected runaway stars compared to model expectations.}


\fundingInfo{Deutsche Forschungsgemeinschaft (DFG), project NE 515/57-1.}

\maketitle

\section{Introduction}\label{sec1}

Runaway stars are characterised by having higher peculiar velocities than typical field stars, $v_{\text{pec}} \gtrsim 20 - 30\,\text{km\,s}^{-1}$ (e.g. \citealt{Tetzlaff2013, Renzo2019}). There are two suggested ejection mechanisms: (i) In the binary supernova ejection scenario (BES), the secondary component from a binary or multiple system is ejected after the primary explodes in a supernova (SN). A binary becomes unbound if more than half of the total system mass gets suddenly ejected during the SN (\textit{Blaauw kick}, \citealt{Blaauw1961}). Systems that do not fulfil this criterion can still get unbound if the newborn neutron star gets a sufficiently high kick from an asymmetric SN (\citealt{Renzo2019}, hereafter Rz19). The ejected companion flies away with its former orbital velocity. The O9.5 star $\zeta$ Oph and PSR B1929+10 were the first pair of a runaway star and a neutron star suggested to have been together as a massive binary \citep{Hoogerwerf2000, Hoogerwerf2001}. However, later works \citep{Chatterjee2004, Tetzlaff2010, Zehe2018} show that the association is very unlikely. Recently, \citet{Neuhaeuser2019} connected $\zeta$ Oph to PSR B1706$-$16. Using Monte Carlo simulations, they traced both objects back to a common origin in Upper-Centaurus-Lupus, a subgroup of the Scorpius-Centaurus-Lupus OB association, $1.78\pm0.21$\,megayears (Myr) ago. The moderate distance $d=107\pm4$\,pc of the corresponding SN makes it likely that this event contributed to the $^{60}$Fe content that was detected in the Earth crust and ocean sediments, dated to an arrival time at the Earth $\sim$2\,Myr ago (\citealt{Knie1999}; \citealt{Breitschwerdt2016}; \citealt{Wallner2016}). (ii) In the dynamical ejection scenario (DES), the runaway star is ejected by gravitational interaction of stars in dense, multiple stellar systems, e.g. globular clusters or OB associations \citep{Poveda1967}. Examples are AE Aur and $\mu$ Col, which were ejected from the Trapezium Cluster in the Orion Nebula, possibly by an encounter with the massive binary $\iota$ Ori \citep{Hoogerwerf2001}. This mechanism generally produces higher velocities than the BES (Rz19), whereas the highest stellar velocities can be explained by the Hills mechanism: The tidal encounter with the supermassive black hole in the Galactic centre \citep{Hills1988} can eject a \textit{hypervelocity star} with up to $v_{\text{pec}} \gtrsim 1000\,\text{km\,s}^{-1}$.

The limit $v_{\text{pec}} \gtrsim 20 - 30\,\text{km\,s}^{-1}$ is based on statistical reasons. \cite{Tetzlaff2013} fitted the velocity distribution of nearby stars with two Maxwellians; one for the regular population and one for the high-velocity population. Both intersect at $v_{\text{pec}} \approx 25\,\text{km\,s}^{-1}$. However, Rz19 performed an extensive population synthesis of massive binaries and emphasised that up to 95\,\% of SN-ejected main-sequence (MS) companions are expected to have $v_{\text{pec}} \leq 30\,\text{km\,s}^{-1}$ and should be called \textit{walkaway stars}. In this work, we will not divide between \textit{runaway} and \textit{walkaway} stars. All stars with a possible BES origin will be classified as runaway star candidates, regardless of their velocity.

While both ejection scenarios are verified by observations, it is still unclear whether one of them is dominant. Therefore, it is important to observe a large number of runaway stars to find whether they are produced by the BES or the DES. One possibility to confirm a BES origin is to look for SN debris in the stellar atmosphere, which can principally be done by high-resolution spectroscopy (e.g. \citealt{Przybilla2008}). Another possibility is to identify runaway stars inside SN remnants (SNRs) or in the vicinity of neutron stars (e.g. \citealt{Dincel2015}). SNRs are only visible for up to $\sim$0.1\,Myr, which constrains the flight time since the SN and therefore the search radius and the maximum age of a runaway star. If the trajectory can be traced back to a common location with the SNR centre and/or the neutron star position in the past, the star should have been ejected during the SN. We then know the time and place of the SN and can constrain SNR expansion models, using the expansion time and the offset between the geometric centre and the actual explosion site. If the age of the runaway star is known, e.g. from a birth cluster, by subtracting the flight time we obtain the lifetime of the SN progenitor and therefore its mass. From the space velocity of the runaway star, we can determine the pre-SN orbital parameters and the SN birth kick velocity imparted on the neutron star (e.g. \citealt{Dincel2015, Neuhaeuser2019}).

Previous searches for runaway stars have mainly focused on OB-type stars. They are not only brighter, but also, due to their limited lifetime, they did not travel too far away from their birth location (up to $\sim$820\,pc for a B9.5 star travelling with $\sim$25\,km\,s$^{-1}$). OB runaway stars were searched for both outside of SNRs (e.g. \citealt{Hoogerwerf2000, Hoogerwerf2001}; \citealt{Tetzlaff2010}; \citealt{Tetzlaff2011}; \citealt{Tetzlaff2013}; \citealt{Tetzlaff2014}), and inside SNRs (e.g. \citealt{Dincel2015}; \citealt{Boubert2017}, hereafter Bo17).

\citet{Dincel2015} found the B0.5\,V runaway star HD\,37424 ($v_{\text{pec}}=74\pm8\,\text{km\,s}^{-1}$) in the SNR S147 and PSR\,J0538+2817 to have been close to the geometric centre of the SNR $30\pm4$\,kyr ago, confirming that the BES does happen. Bo17 confirmed HD\,37424 and suggested three further likely candidates, located in the SNRs Cygnus Loop, HB21 and Monoceros Loop. They did not exclude late-type runaways, but they considered early-type stars to be more probable.

For our search in nearby SNRs, we know that an ejected companion has to be young. A SN progenitor has a MS lifetime of up to 32\,Myr (\citealt[Table 3]{Ekstroem2012}, in the case of an $8\,M_{\odot}$ star). The timespans of the subsequent burning cycles and the lifetime of the SNR ($\sim$0.1\,Myr) are negligible compared to that. For dwarf stars of spectral types mid-F to M, we can use lithium (Li\,I) as youth indicator, because it gets depleted in these stars during the early phases of stellar evolution (e.g. \citealt{Bodenheimer1965}; \citealt{Dantona1984}; \citealt{Covino1997}; \citealt{Neuhaeuser1997}); we use spectra of these stars to search for the absorption line at 6707.8\,\AA{}. We can estimate the age by measuring the effective temperature of the stars and the equivalent width of the Li 6708\,\AA{} line and comparing it to stars in clusters with known ages, e.g. the Pleiades (\citealt{Soderblom1993}; \citealt{Soderblom2010a, Soderblom2010b}).

In this work, we present results of our search for runaway stars of all spectral types in twelve nearby SNRs with data from the \textit{Gaia} mission \citep{Gaia2016b}. In Section~\ref{sec2} we explain how the SNRs and corresponding runaway star candidates were selected. In Section~\ref{sec3} we describe the observations. In Section~\ref{sec4} we show the spectral analysis and estimate the ages of our candidates. In Section~\ref{sec5} we discuss the results and possible implications for the frequency of runaway stars from the BES and, finally, in Section~\ref{sec6} we conclude our work and give an outlook on future work.

\section{Candidate selection}\label{sec2}

\subsection{Selection of SNRs}\label{sec2.1}

Two main resources were used to find suitable SNRs, namely (i) the Green catalogue of SNRs\footnote{\href{http://www.mrao.cam.ac.uk/surveys/snrs/}{http://www.mrao.cam.ac.uk/surveys/snrs/}} \citep{Green2014} for positions and sizes and (ii) the catalogue from the University of Manitoba\footnote{\href{http://www.physics.umanitoba.ca/snr/SNRcat}{http://www.physics.umanitoba.ca/snr/SNRcat}} (\citealt{Ferrand2012} and references therein), compiled mainly from high-energy observations, for distances and ages.

The project was started in 2016 and the first target selection was done before \textit{Gaia} data release (DR) 2 \citep{Gaia2018} was available. Therefore, \textit{Gaia} DR1 \citep{Gaia2016a} was used for these targets. The much larger number of stars in \textit{Gaia} DR2 had consequences for the SNR selection, which are described in the following.

For observations of late-type stars, it was necessary to set a magnitude limit and, hence, a distance limit for the SNRs. While the limiting magnitude of a spectrograph at an 8\,m-class telescope is $m_{V,\text{max}}\approx19.5$\,mag (see UVES/VLT manual), for our selection we use $m_G\leq17.0$\,mag in order to reach a sufficient signal-to-noise ratio, $S/N\gtrsim30$, in the spectra (see Section~\ref{sec3}). Using this limit, mass-brightness relations for the main-sequence from \cite{Henry1993}, and choosing a limiting distance of $d<500$\,pc, where four SNRs are available, we get a mass limit of $M=0.60\,M_{\odot}$, corresponding to spectral type K8. These limits were applied for stars selected from \textit{Gaia} DR2. For stars selected earlier from \textit{Gaia} DR1, \textit{Tycho-Gaia Astrometric Solution} (TGAS), we chose $d<1600$\,pc, including up to 25 SNRs. With this limit we get down to $M=0.86\,M_{\odot}$, corresponding to spectral type K0.5. The limits were chosen as a compromise between investigating a high number of SNRs and getting down to the latest possible spectral types. Due to the higher number of stars in \textit{Gaia} DR2, we limited the distance to $d<500$\,pc in order to create feasible projects for follow-up observations.

Analysing historical SNRs is particularly interesting, because their age is accurately known, so we can directly study the SNR expansion and if we find a runaway star, the determination of the distance and pre-SN binary properties can be done much more accurately. Therefore, we also added four historical SNRs to our sample, despite their larger distances. For the most distant one, Cassiopeia A at $d\approx3.5$\,kpc, with the limits described above we could detect stars down to spectral type $\sim$F9. The limiting spectral types of the other SNRs, as well as further properties, are given in Table~\ref{tab1}\hspace{-1.7mm}, where we list our selection of the four SNRs with $d\leq500$\,pc, four further SNRs with $500\text{\,pc}\leq d\leq1600$\,pc and the four historical SNRs Cassiopeia\,A, 3C58 (SN\,1181), Crab Nebula (SN\,1054) and SNR G347.3$-$00.5 (SN\,393).

\begin{table*}%
\centering
\caption{Properties of the SNRs studied in this work. The equatorial coordinates RA/DEC and angular diameters $\Theta$ are taken from \citet{Green2014}, the distances $d$ and ages $t$ from \citet{Ferrand2012} and references therein, where the errors come from the range between the lowest and highest values in the given literature. For 3C58, Crab Nebula and SNR G347.3$-$00.5 the ages are accurately known (assuming that the associations to the historical SNe are correct). The last column gives the limiting stellar spectral type SpT$_{\text{lim}}$ that could be observed at the corresponding nominal distance and a limiting magnitude of $m_G=17.0$\,mag, as descibed above.}
\label{tab1}
\begin{tabular*}{500pt}{cccccccc}
\hline
SNR Name & Alternative Name & RA [h:m:s] & DEC [d:m] & $\Theta$ [arcmin] & $d$ [kpc] & $t$ [kyr] & SpT$_{\text{lim}}$ \\
\hline
G074.0$-$08.5 & Cygnus Loop        & 20:51:00 & $+$30:40 & $230\times160$ & $0.79\pm0.21$ & $15\pm5$ \hspace{1mm} & K5  \\
G160.9$+$02.6 & HB9                & 05:01:00 & $+$46:40 & $140\times120$ & $0.80\pm0.40$ & $5.5\pm1.5$ & K5  \\
G180.0$-$01.7 & S147               & 05:39:00 & $+$27:50 & $180\times180$ & $1.30\pm0.20$ & $30\pm4$ \hspace{1mm} & K1.5 \\
G205.5$+$00.5 & Monoceros Loop     & 06:39:00 & $+$06:30 & $220\times220$ & $1.44\pm0.54$ & $90\pm60$ & K1.5  \\
G260.4$-$03.4 & Puppis A           & 08:22:10 & $-$43:00 & $60\times50$ & $1.3\pm0.3$ & $4.08\pm0.38$ & K1.5 \\
G263.9$-$03.3 & Vela               & 08:34:00 & $-$45:50 & $255\times255$ & $0.275\pm0.025$ & $18\pm9$ \hspace{1mm} & M0.5  \\
G266.2$-$01.2 & Vela Jr.           & 08:52:00 & $-$46:20 & $120\times120$ & $0.75\pm0.25$ & $3.8\pm1.4$ & K5  \\
G330.0$+$15.0 & Lupus Loop         & 15:10:00 & $-$40:00 & $180\times180$ & $0.33\pm0.18$ & $23\pm8$ \hspace{1mm} & M0.5  \\
G111.7$-$02.1$^*$ & Cassiopeia A   & 23:23:26 & $+$58:48 & $5\times5$ & $3.50\pm0.20$ & $0.334\pm0.018$ & F9 \\
G130.7$+$03.1$^*$ & 3C58 (SN 1181) & 02:05:41 & $+$64:49 & $9\times5$ & $2.60\pm0.60$ & 0.839 & G2  \\
G184.6$-$05.8$^*$ & Crab Nebula    & 05:34:31 & $+$22:01 & $7\times5$ & $1.85\pm0.35$ & 0.966 & G9  \\
G347.3$-$00.5$^*$ & SN 393         & 17:13:50 & $-$39:45 & $65\times55$ & $1.3\pm0.4$ & 1.627 & K1.5  \\
\hline
\end{tabular*}
\begin{tablenotes}
\item $^*$ Historical SNR.
\end{tablenotes}
\end{table*}

\subsection{Selection of runaway star candidates}\label{sec2.2}

\begin{figure}
\includegraphics[width=84mm]{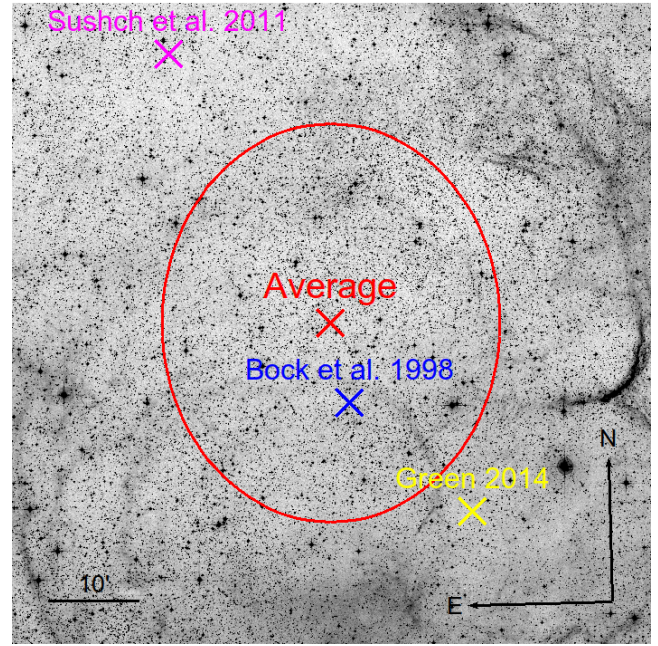}
\caption{The central $\ang{;70;} \times \ang{;70;}$ of the Vela SNR. The crosses mark the three geometric centres from X-rays (magenta, \citealt{Sushch2011}), radio (blue, \citealt{Bock1998}) and the Green catalogue (yellow, \citealt{Green2014}). Their average and standard deviation are shown by a red cross and ellipse, respectively. Background image from ESO DSS-2-red.\label{Velacenters}}
\end{figure}

In order to define an area for the search for runaway star candidates, it is important to know the position of the SN, which should be close to the geometric centre (GC) of the SNR. In most cases, we chose the coordinates from \cite{Green2014}, which are mainly based on radio observations, where the SNR morphology usually is best visible.
In the case of the Vela SNR, we combined the centre from \cite{Green2014} with two images from different wavelengths, namely the 843\,MHz radio image from \cite{Bock1998} and the X-ray image from \cite{Sushch2011}. The average coordinates of the three solutions were taken as the actual GC (see Fig.~\ref{Velacenters}\hspace{-1.7mm})\footnote{Note that in the case of the Vela SNR we did not trace back the stellar trajectories to the GC, but to the past position of the Vela pulsar.}.

Around the centre, we defined a search radius, depending on the maximum possible velocity of runaway stars and the age of the SNR. \citet{Tauris2015} gives a maximum ejection velocity of $v_{\text{max}}=1050\,\text{km\,s}^{-1}$ for stars with $0.9\,M_{\odot}$, corresponding to extreme cases which are very rare (Rz19). We use $v_{\text{max}}=1000\,\text{km\,s}^{-1}$, which yields a search radius of

\begin{equation}
    r_{\text{search}} = 3.4934 \times \frac{t\,\text{[yr]}}{d\,\text{[pc]}}\,\text{arcmin},
	\label{rsearch}
\end{equation}

with the age $t$ and the distance $d$ of the SNR. As SNR expansion velocities are well above typical runaway star velocities during the \textit{Sedov-Taylor} phase, which is the current state of most of our SNRs, we expect potential runaway stars to still be located inside the SNR.

For each SNR, we first selected all \textit{Gaia} stars in the given search radius with $m_G=G\leq17$\,mag. We then traced back the projected trajectories of the stars, using their proper motions and the age of the SNR to obtain their coordinates at the time of the SN. We neglected the Galactic gravitational potential because the lifetime of a SNR is so short that the potential will not have a significant effect. Gaussian error propagation was used to calculate the uncertainties of the past positions.
We selected the stars that, at the time of the SN, were located inside the error ellipse of the GC. For stars from \textit{Gaia} DR1, we used the standard deviations in right ascension and declination from the determination of the Vela GC to define the error ellipse, which was then also used for the other SNRs, scaled to the corresponding diameter as given in \citet{Green2014}. For stars from \textit{Gaia} DR2, we used the error estimate given by \citet{Green2009}, which we translate here to

\begin{eqnarray}
\Delta \text{GC} &=& \ang{0.021;;} \times \ang{0.017;;}, \text{~if~} \ang{;0;} \leq \Theta \leq \ang{;50;} \nonumber \\
\Delta \text{GC} &=& \ang{0.042;;} \times \ang{0.034;;}, \text{~if~} \ang{;50;} \leq \Theta \leq \ang{;100;} \label{SNRerror}  \\
\Delta \text{GC} &=& \ang{0.063;;} \times \ang{0.051;;}, \text{~if~} \ang{;100;} \leq \Theta  \nonumber
\end{eqnarray}

where $\Theta$ is the angular diameter of the SNR. However, if a neutron star is associated with the SNR, we selected only the stars that could be traced back to a possible common origin with the neutron star. This was the case for S147, Vela, 3C58 and the Crab Nebula. Questionable cases or neutron stars with unknown proper motions were not considered.

Note that the strict limitation of Eqn.~\ref{SNRerror} means that we loose up to $\sim$33\,\% of runaway star candidates that were located more than $1\,\sigma$ from the nominal GC. We decided to make this limitation in order to create feasible observing projects, concentrating on the most promising candidates.

\begin{figure*}
	\centerline{\includegraphics[width=170mm
	]{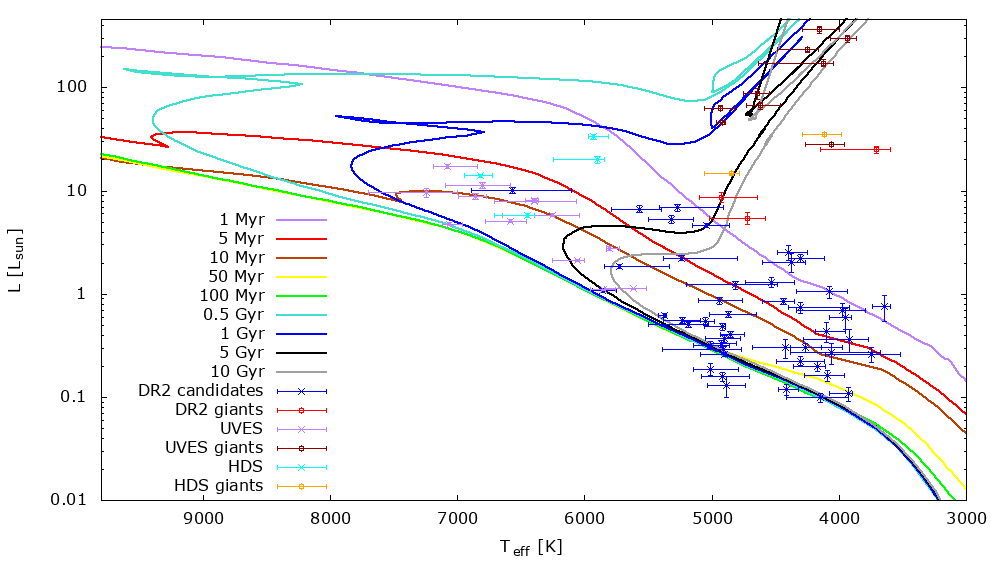}}
	\caption{HRD of the runaway star candidates with effective temperatures and luminosities taken from \textit{Gaia} DR2. In the key, for stars that were observed by us, we give the names of the corresponding spectrographs, UVES and HDS. For comparison, we show isochrones calculated from PARSEC models \citep{Bressan2012}. The stars marked in red, dark-red and orange were considered to be post-MS giants and ruled out from the investigation.}
	\label{HRDL}
\end{figure*}

We also checked if the \textit{Gaia} parallaxes of the stars are consistent with the range of possible distances of the SNR. We compared these to distances derived by \citet{Bailer-Jones2018} and found no significant deviations for the parallax range considered in our work. All candidates are still within the correct range when using their distances from \citet{Bailer-Jones2018}. For stars from \textit{Gaia} DR2, we also checked the given values for radius, luminosity, effective temperature and surface gravity, so that giants could be excluded beforehand. The selection could be made by plotting the candidates into a Hertzsprung-Russell diagram (HRD), shown in Fig.~\ref{HRDL}\hspace{-1.7mm}, where we compare our candidates to PARSEC isochrones\footnote{PAdova and TRieste Stellar Evolution Code, see \href{http://stev.oapd.inaf.it/cgi-bin/cmd_3.3}{http://stev.oapd.inaf.it/cgi-bin/cmd\_3.3}} \citep{Bressan2012}. Reddish colours indicate if a star was classified as a giant. If a star has an increased luminosity compared to the zero-age main-sequence (ZAMS), it can be either a pre-MS star or an evolved star (i.e. terminal-age MS or giant). As we are looking for young stars, it might be better to leave a few evolved stars in the sample than excluding too many pre-MS stars. The division was roughly made at the isochrone for 1\,Myr. For questionable cases and because luminosities were not given for all possible targets in \textit{Gaia} DR2, we also checked the surface gravities $\log(g)$ of the targets in the \textit{StarHorse} catalogue \citep{Anders2019}. Four targets with $\log(g)<3$, where no luminosity was given, were excluded.

\begin{figure}
	\centerline{\includegraphics[width=84mm
	]{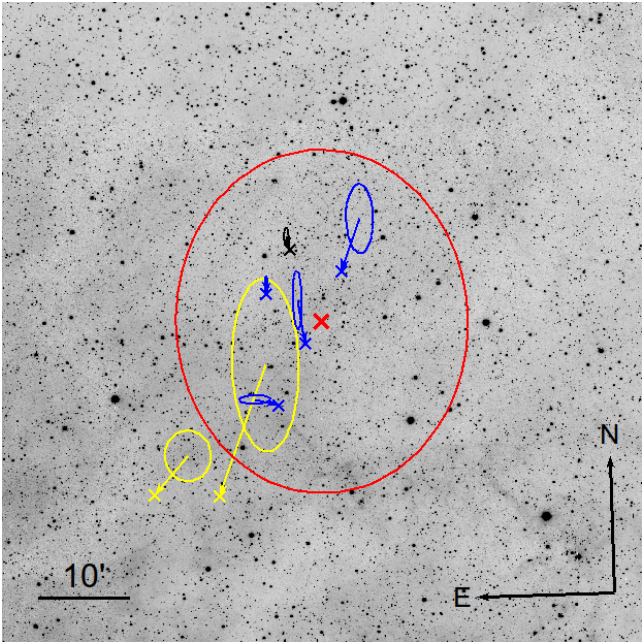}}
	\caption{The runaway star candidates in the Monoceros Loop. The image shows the central $\ang{;70;} \times \ang{;70;}$ of the SNR. The geometric centre of the SNR is marked by a red cross with corresponding $1 \sigma$ error ellipse. The current positions of runaway star candidates with and without detected lithium are marked by yellow and blue crosses, respectively. They are connected with the positions at the time of the SN (surrounded by a corresponding error ellipse) by arrows in the direction of their proper motion. The star marked in black is HD\,261393, which was suggested as a runaway star by Bo17 and also observed by us. The background image was taken by us in H$\alpha$ with the \textit{Schmidt-Teleskop-Kamera} \citep{Mugrauer2010} at the University Observatory Jena.}
	\label{Mono}
\end{figure}

Stars that fulfill all criteria were considered as good runaway star candidates and selected as targets for the spectroscopic follow-up observations. In Fig.~\ref{Mono}\hspace{-1.7mm} the selection is shown for the Monoceros Loop. Among the 21 stars found here from \textit{Gaia} DR1, we only show the remaining runaway star candidates and two stars that show Li in their spectra. No neutron star is known in the Monoceros Loop, so the GC was used as a reference position.

After the detection and characterisation of runaway star candidates, the goal was to determine the radial velocity (RV) and the atmospheric parameters and to search for the Li 6708\,\AA{} line as youth indicator in the spectra of late-type stars.

\section{Observations}\label{sec3}

\subsection{VLT UVES Spectroscopy}\label{ObsUVES}

For southern targets, we have used the \textit{UV-Visual Echelle Spectrograph} (UVES), which is mounted at the \textit{Nasmyth} focus of \textit{Kueyen} (UT2) at the \textit{Very Large Telescope} (VLT), run by the \textit{European Southern Observatory} (ESO) and located on the \textit{Cerro Paranal} in Chile. We observed 33 runaway star candidates, selected from \textit{Gaia} DR1 TGAS, in five SNRs on the southern sky, namely (i) G180.0$-$01.7 (S147), (ii) G205.5+00.5 (Monoceros Loop), (iii) G260.4$-$03.4 (Puppis A), (iv) G263.9$-$03.3 (Vela) and (v) G266.2$-$01.2 (Vela Jr.).

We used the Red arm (Ra), cross-disperser 3, with a central wavelength of 600\,nm. In this setup, two charge-coupled device (CCD) detectors are used which cover the spectral ranges $4986-5957$\,\AA{} and $6036-7003$\,\AA{}. Using $1\times1$ binning, which is adequate for the relatively bright targets observed in P100, a slit width of $\ang{;;1.2}$ yields a resolving power of $R=32250$.

For the used magnitude limit of $G=17$\,mag, corresponding to $V=17.13$\,mag\footnote{For the calculation, $G=17.00$\,mag was converted to $V=17.13$\,mag by using the polynomial relations by \citet{Evans2018} with $BP-RP=0.782$ given in \citet[Table 5]{Pecaut2013} for a G0\,V star.}, with the given setup we reach $S/N=33$ at 6711\,\AA{} in one hour integration time, derived with the UVES exposure time calculator (ETC), version P106.2\footnote{\href{https://www.eso.org/observing/etc/bin/gen/form?INS.NAME=UVES+INS.MODE=spectro}{https://www.eso.org/observing/etc/bin/gen/form?INS.NAME$=$UVES$+$\\INS.MODE$=$spectro}}, for a G0\,V star at airmass 1.1, with a seeing of $\ang{;;1}$ and with $2\times2$ binning. A G0\,V star was chosen for this example calculation because it represents a typical target of our selection.

The VLT observations were executed between October and December 2017 in service mode. In the appendix, Table~\ref{Obslist}, we list the 33 runaway star candidates observed with UVES. The individual exposure times varied between 10\,s and 500\,s, depending on the brightness of the star. Two exposures were taken for each star, where an average $S/N$ of 64.7 was achieved for the single exposures. The total on-source integration time was 5 hours and 4 minutes.

Standard calibrations were used, i.e. Bias frames, Flatfield images of a Halogen lamp and wavelength calibration images of a Thorium-Argon (ThAr) lamp which are taken regularly for each standard setup as described in the UVES calibration plan. Data reduction was done with the \textit{EsoReflex} pipeline for UVES. The individual reduction steps are bias subtraction, order detection, flatfielding, wavelength calibration and spectrum extraction. We obtained four spectra for each star; for each of the two exposures, we got one spectrum for the lower and one for the upper wavelength regime.

The two exposures for each star were averaged to one spectrum with \textit{IRAF}. Normalisation was done with \textit{iSpec} \citep{Blanco-Cuaresma2014}.

\subsection{Subaru HDS Spectroscopy}\label{ObsHDS}

Six northern targets, all located in SNR G160.9+02.6 (HB9), were observed with the High Dispersion Spectrograph (HDS), mounted at the \textit{Nasmyth} focus of the Subaru telescope, which is run by the \textit{National Astronomical Observatory of Japan} (NAOJ) and located on the \textit{Mauna Kea}, Hawaii.

HDS uses two CCD detectors, where with the Ra setup the first one covers the wavelength range $5062-6446$\,\AA{} and the second one $6509-7890$\,\AA{}. We used $2\times2$ binning and a slit width of $\ang{;;1.0}$, yielding a resolving power of $R=36000$.

For this setup, with the HDS ETC\footnote{\href{https://www.naoj.org/cgi-bin/hds_etc.cgi}{https://www.naoj.org/cgi-bin/hds\_etc.cgi}}, we obtain $S/N\approx24$ at 6697\,\AA{} for a G0\,V star with $V=17.13$\,mag at airmass 1.1, with a seeing of $\ang{;;1}$ and an exposure time of one hour.

At the bottom of Table~\ref{Obslist}\hspace{-1.7mm} we list the six runaway star candidates that were observed with Subaru in period S18B. All observations were done on 2018 Oct 26. The individual exposure times varied between 40\,s and 300\,s. Two exposures were taken for each target and $S/N=105.7$ was reached on average for the single exposures.

Standard calibrations were used and data reduction was done manually with \textit{IRAF}. We corrected for the bias level and bad pixels and applied order tracing, flatfielding, spectrum extraction and wavelength calibration, following the procedure described in \cite{Aoki2014}. As described for UVES/VLT, also here we obtained four spectra for each star. The last step of the data reduction already included the normalisation of the spectra by fitting a spline to the continuum, before the spectral orders were merged.

\section{Spectral Analysis}\label{sec4}

\begin{figure*}
	\centerline{\includegraphics[width=155mm
	]{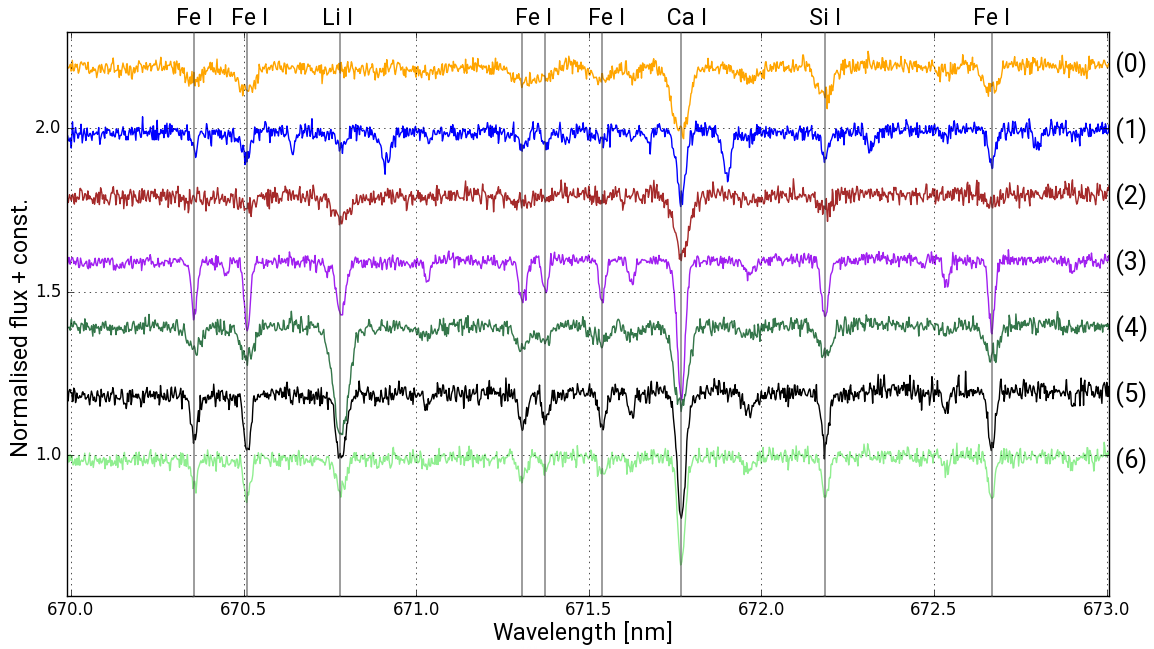}}
	\caption{Normalised, RV corrected and shifted UVES/VLT spectra between $\sim$6700\,\AA{} and 6730\,\AA{}. The bottom six spectra show our Li-rich targets, which have spectral types between F2 and G2, in particular (1) TYC159-251-1, (2) TYC159-343-1, (3) TYC8150-2802-1, (4) TYC8150-3105-1, (5) TYC8152-1456-1 and (6) TYC8152-550-1. The yellow spectrum (0) shows TYC159-1896-1, a mid-F-type star without significant Li absorption, for comparison. The varying strengths of the Li 6707.8\,\AA{} line can be seen by comparison to the strong, nearby lines: Fe\,I at 6704.5\,\AA{}, 6705.1\,\AA{}, 6713.1\,\AA{}, 6713.8\,\AA{} and 6726.7\,\AA{}, Ca\,I at 6717.7\,\AA{} and Si\,I at 6722.0\,\AA{}. The individual spectral types and Li equivalent widths are displayed in Table~\ref{Li}\hspace{-1.7mm}.}
	\label{comp_Li}
\end{figure*}

In the averaged spectra, we searched for the Li 6708\,\AA{} absorption line to measure the equivalent width (see Section~\ref{LiEW}). Li was found in ten of our VLT spectra. Four of these stars were found to be giants (see Fig.~\ref{HRDL}\hspace{-1.7mm}) and are therefore too old to be runaway stars. Fig.~\ref{comp_Li}\hspace{-1.7mm} shows seven of our VLT spectra around the Li line. While TYC159-1896-1 is only shown for comparison, six of them show Li absorption in their spectra, namely (1) TYC159-251-1 and (2) TYC159-343-1, located in the Monoceros Loop, (3) TYC8150-2802-1 and (4) TYC8150-3105-1 (Vela) as well as (5) TYC8152-1456-1 and (6) TYC8152-550-1 (Vela Jr.). These six stars were analysed further with \textit{iSpec} \citep{Blanco-Cuaresma2014} to determine their RVs and atmospheric parameters.

\subsection{A new spectroscopic binary}

\begin{figure*}
	\centerline{\includegraphics[width=170mm
	]{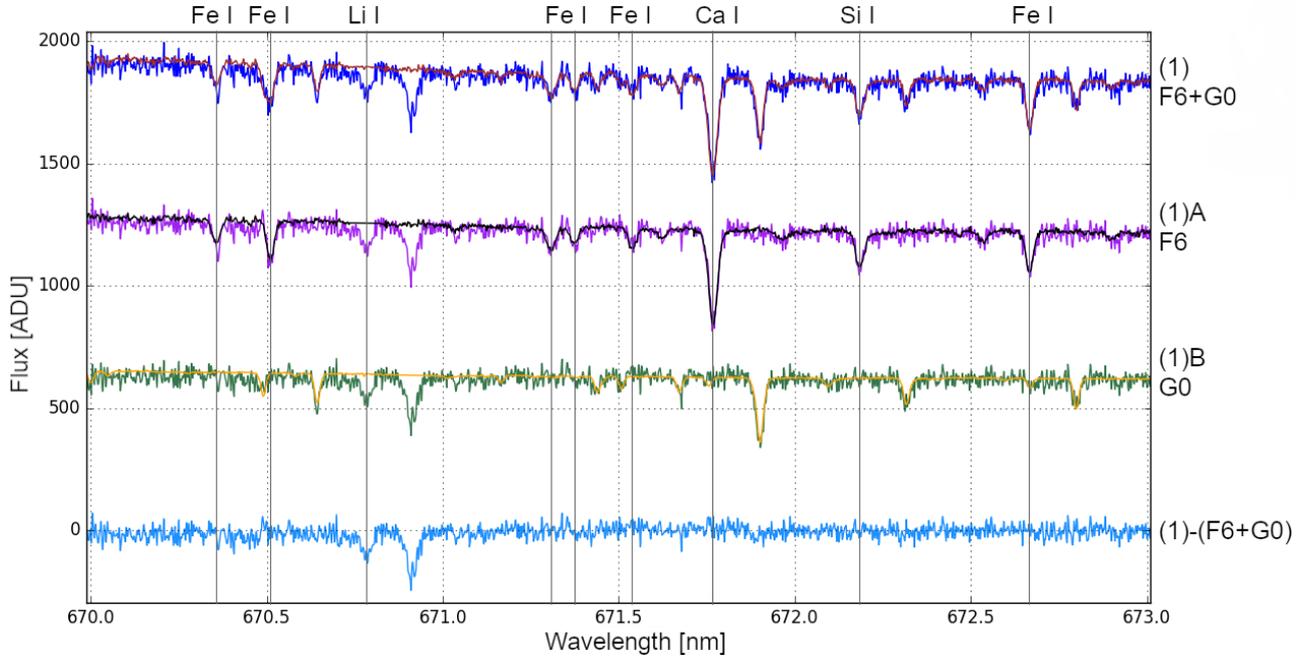}}
	\caption{The spectrum of TYC159-251-1 (blue, 1) and its two components TYC159-251-1\,A (purple, 1A) and TYC159-251-1\,B (green, 1B) between $\sim$6700\,\AA{} and 6730\,\AA{}. The first component was used for RV correction. For comparison we also show the F6 template spectrum (black), scaled by the factor 0.661, and the G0 template (yellow), scaled by the factor 0.339, which were subtracted from the observed spectrum to obtain the two components. The red spectrum is the addition of the two template spectra, while the residuals between the observed spectrum and the added templates are shown in light-blue. Lithium was manually removed from the template spectra.}
	\label{TYC159-251-1}
\end{figure*}

The spectrum of TYC159-251-1 shows additional absorption lines, each redshifted by $\sim$1.3\,\AA{} compared to the main absorption lines. Therefore, the star can be identified as a double-lined spectroscopic binary (SB2). The system is not yet listed in the 9th Catalogue of Spectroscopic Binary Orbits (SB9, \citealt{Pourbaix2009}). In the following, we will use the designation TYC159-251-1\,A for the primary (brighter) component, TYC159-251-1\,B for the secondary component and TYC159-251-1 for the whole system. As we would get misleading results from analysing the observed spectrum which contains light from both stars, we disentangled the two components by subtracting a G0 template spectrum from the observed spectrum to obtain the first component and a F6 template to obtain the second component (both templates taken from \citealt{Bagnulo2003})\footnote{The spectral types F6 and G0 were found to yield the best representation of the observed spectrum after iterating with different combinations of template spectra from \citet{Bagnulo2003} between F6 and G0 (F6+F9, F6+G0, F7+F9, F8+F8, F8+G0 and F9+F9. Other combinations were excluded beforehand).}. The continuum levels of the templates were normalised before to have the flux level of the combined spectrum of TYC159-251-1, multiplied by the factors 0.661 and 0.339, respectively, for the two components, corresponding to the luminosities of the used spectral types according to \citet[Table 5, hereafter PM13]{Pecaut2013}. Adding up these modified templates yielded the best representation of the observed spectrum. Fig.~\ref{TYC159-251-1}\hspace{-1.7mm} shows the observed spectrum (blue) as well as the individual components (magenta, green) after subtracting the template spectra (black, yellow). The residuals (light-blue) between the observed spectrum and the added templates (red) only show a significant deviation from zero at the Li lines, which is due to the manual removal of Li in the templates (to retain it in the component spectra). In the following subsections, potential peculiarities in the analysis of this SB2 will be described in additional paragraphs.

\subsection{Radial and space velocities}
\label{sec4RV}

The barycentric correction was done with \textit{iSpec} before comparing the spectral lines to an atomic line mask (cross-correlation) to determine the RVs and shift the spectra correspondingly. For the cross-correlation, a line list from a \textit{Narval} solar spectrum ($370-1048$\,nm) was used. For the early-type stars, the RV determination with \textit{iSpec} was not possible due to the fewer and broader lines. We determined their RVs manually with \textit{IRAF} from the positions of the following lines, if available: Fe\,II 5018.4\,\AA{}, Mg\,I 5172.7\,\AA{}, Mg\,I 5183.6\,\AA{}, He\,I 5875.6\,\AA{}, Si\,II 6347.1\,\AA{}, Si\,II 6371.4\,\AA{}, Fe\,II 6456.4\,\AA{}, H$\alpha$ 6562.8\,\AA{}. The central wavelength for each line was determined as the average centre from three Gaussian fits, whereupon the $v_r$ for each line were calculated. The $v_r$ in Table~\ref{veloc}\hspace{-1.7mm} correspond to average and standard deviation of these measurements.
By combining the RVs with proper motions and parallaxes from \textit{Gaia} DR2, we calculated the heliocentric space velocities of the stars. For the calculation of the peculiar velocity with respect to the local standard of rest, we first calculated the Galactic space velocity components $(U,V,W)$, following the equations given in \citet{Johnson1987}. Then we corrected for the solar motion with respect to the local standard of rest by adding $(U_{\odot},V_{\odot},W_{\odot})=(8.5,13.38,6.49)\,\text{km\,s}^{-1}$ \citep{Coskunoglu2011}. The peculiar velocity is the absolute value of the resulting vector. The kinematic data of the six Li-rich stars and the best early-type candidates are given in Table~\ref{veloc}\hspace{-1.7mm}. TYC8150-2802-1, TYC8152-1456-1, TYC8152-550-1, TYC3344-235-1, TYC159-2771-1 and HD\,76060 have $v_{\text{pec}}>25\,\text{km\,s}^{-1}$ and could therefore be classified as runaway stars according to \cite{Tetzlaff2013}.

For the SB2 system TYC159-251-1, we determined the momentary RVs of the two components with \textit{IRAF}, measuring the positions of 38 absorption lines of Fe\,I, Fe\,II, Ca\,I, Ni\,I and Si\,II. As line list we chose the \textit{Identification List of Lines in Stellar Spectra} (ILLSS, \citealt{Coluzzi1999}). To obtain the systemic RV $\gamma$, we need the mass ratio $q=M_2/M_1<1$. In Section~\ref{atm_params} we will find the components to have spectral types F6\,$-$\,F7 and F9\,$-$\,G0, respectively, corresponding to $q=0.91^{+0.03}_{-0.05}$. The systemic RV can be calculated as

\begin{equation}
\gamma = v_{r_1} + \frac{q}{1+q} \times (v_{r_2}-v_{r_1})
\end{equation}

The following results were obtained:

\begin{equation}
\begin{split}
v_{r_1} &= -10.7\pm0.6\,\text{km\,s}^{-1} \\
v_{r_2} &= 48.4\pm0.7\,\text{km\,s}^{-1} \\
\gamma &= 17.5^{+0.9}_{-1.1}\,\text{km\,s}^{-1} \\
\end{split}
\end{equation}

The systemic RV is then used to determine the space motion (see $v_{\text{pec}}$ in Table~\ref{veloc}\hspace{-1.7mm}). Extensive follow-up RV monitoring is required to fit a RV curve and determine the orbital parameters. Only then the systemic and space velocity of the system can be given reliably.

\begin{table*}
\caption{\textit{Gaia} DR2 $G$ magnitudes, parallaxes $\pi$ and kinematic parameters of the six Li-rich stars as well as the remaining early-type candidates in HB9, Monoceros Loop and Vela Jr. Parallaxes and proper motions are from \textit{Gaia} DR2, the barycentric corrections ($BC$) were determined with \textit{iSpec}. The radial velocities $v_r$ of the Li-rich stars were determined with \textit{iSpec}, the $v_r$ of the early-type stars were determined with \textit{IRAF}. For TYC159-251-1, we give the systemic RV $\gamma$ of the binary, approximated using a mass ratio of $q=0.91^{+0.03}_{-0.05}$. The individual RVs of the binary components were derived with \textit{IRAF}. For the spectral types, see Table~\ref{Li}\hspace{-1.7mm}.}
\label{veloc}
\centering
\begin{tabular*}{510pt}{cccccccc}
\hline
Name & $G$ & $\pi$ & $\mu_{RA}$ & $\mu_{DEC}$ & $BC$ & $v_r$ & $v_{\text{pec}}$ \\
 & [mag] & [mas] & [mas\,yr$^{-1}$] & [mas\,yr$^{-1}$] & [km\,s$^{-1}$] & [km\,s$^{-1}$] & [km\,s$^{-1}$] \\
\hline
Li-rich stars & & & & & & & \\
TYC159-251-1   & 11.51 & $1.774\pm0.041$ & $+3.63\pm0.07$ & $-9.44\pm0.06$  & 5.98 & $+17.5^{+0.9}_{-1.1}$ & $19.3^{+1.1}_{-1.2}$ \\
TYC159-343-1   & 10.87 & $1.383\pm0.042$ & $+2.51\pm0.09$ & $-2.76\pm0.08$   & 6.03 & $+22.27\pm0.41$ & $13.8\pm0.7$ \\
TYC8150-2802-1 & 12.03 & $3.226\pm0.028$ & $-16.94\pm0.05$ & $+29.17\pm0.05$ \hspace{1mm} & 15.14 & $+53.84\pm0.07$ & $58.8\pm0.6$ \\
TYC8150-3105-1 & 11.94 & $3.351\pm0.028$ & $-16.00\pm0.05$ & $+10.05\pm0.04$ \hspace{1mm} & 15.01 & $+20.37\pm0.29$ & $19.7\pm0.5$ \\
TYC8152-1456-1 & 12.47 & $1.661\pm0.037$ & $+5.17\pm0.06$ & $-11.56\pm0.07$ \hspace{1mm}  & 15.13 & $-10.59\pm0.08$ & $49.1\pm1.0$ \\
TYC8152-550-1  & 11.92 & $2.435\pm0.026$ & $+4.95\pm0.04$ & $-10.71\pm0.05$ \hspace{1mm}  & 15.14 & $+36.12\pm0.10$ & $37.1\pm0.5$ \\
\hline
HB9 & & & & & & & \\
TYC3344-235-1 & 11.07 & $0.91\pm0.04$ & $-0.23\pm0.09$ & $+1.37\pm0.06$  & 19.15 & $-31.2\pm2.8$ & $36.0\pm2.9$ \\
TYC3344-679-1 & 11.40 & $1.01\pm0.06$ & $+0.74\pm0.18$  & $-4.76\pm0.11$ & 19.18 & \hspace{1mm} $+8.6\pm7.3$  & \hspace{1mm} $8.2\pm5.9$ \\
TYC3344-683-1 & 12.33 & $1.21\pm0.04$ & $-1.09\pm0.07$ & $-7.02\pm0.05$ & 19.19 & \hspace{1mm} $+9.3\pm7.3$  & $15.3\pm3.6$ \\
TYC3344-553-1 & 10.87 & $1.53\pm0.04$ & $-1.98\pm0.08$ & $-7.22\pm0.06$ & 19.18 & $-14.9\pm7.3$ & $21.9\pm5.7$ \\
\hline
Monoceros Loop & & & & & & & \\
TYC159-2771-1 & 12.01 & $0.75\pm0.04$ & $+1.46\pm0.07$  & $-3.78\pm0.07$ & 19.53 & $+31.8\pm2.0$ & $25.3\pm2.0$ \\
HD261359      & 11.79 & $0.65\pm0.05$ & $-0.48\pm0.08$ & $-3.15\pm0.07$ & 22.59 & \hspace{1mm} $+0.9\pm5.1$  & $17.8\pm4.5$ \\
HD261393      & 10.05 & $0.79\pm0.05$ & $-0.20\pm0.09$ & $-0.97\pm0.07$ & 24.55 & $+31.3\pm5.8$ & $18.5\pm5.3$ \\
TYC159-2337-1 & 11.80 & $0.49\pm0.05$ & $-1.70\pm0.09$ & $-0.45\pm0.08$ & 5.94 &  $+22.1\pm4.7$ & $18.0\pm3.7$ \\
TYC159-2671-1 & 12.25 & $0.69\pm0.04$ & $+0.08\pm0.07$  & $-0.78\pm0.06$ & 19.57 & $+23.3\pm7.3$ & $11.4\pm6.3$ \\
\hline
Vela Jr. & & & & & & & \\
HD\,76060 & 7.85 & $2.06\pm0.04$ & $-12.42\pm0.08$ & $+10.31\pm0.09$ & 9.38 & $+25.7\pm7.3$ & $31.2\pm2.7$ \\
\hline
\end{tabular*}
\end{table*}

\subsection{Atmospheric parameters}\label{atm_params}

With the synthetic spectral fitting technique provided by \textit{iSpec}, we determined the atmospheric parameters of the six Li-rich stars. The routine computes synthetic spectra and compares them to the observed spectra by nonlinear least-squares fitting, minimising the $\chi^2$. We used the continuum-normalised and RV-corrected spectra. The routine considers the following parameters: (i) Effective temperature $T_{\text{eff}}$, (ii) Surface gravity $\log(g)$, (iii) Metallicity [M/H], (iv) Microturbulence velocity $v_t^{\text{mic}}$, (v) Macroturbulence velocity $v_t^{\text{mac}}$, (vi) Rotation $v_{\text{rot}} \sin(i)$ and (vii) Limb darkening coefficient, taking into account the resolving power $R=32250$.

\begin{figure*}
	\centerline{\includegraphics[width=160mm
	]{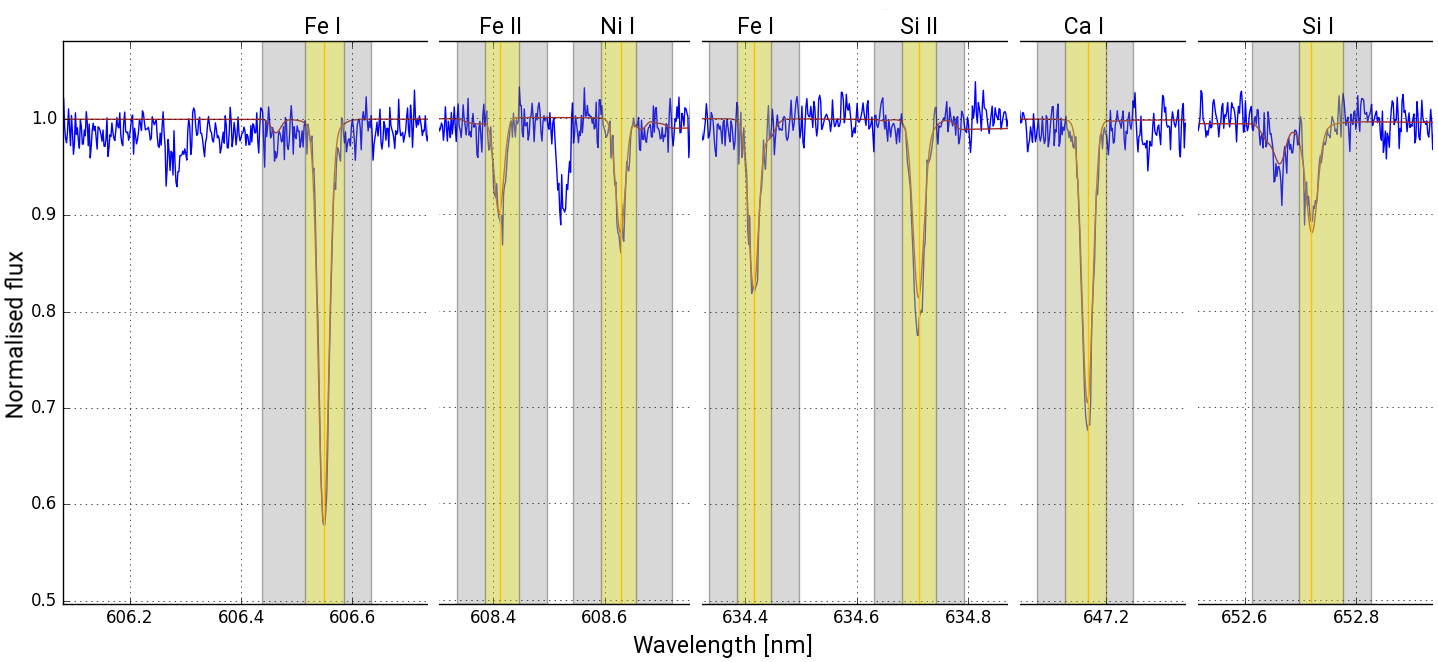}}
	\caption{Different example lines of TYC8152-550-1 (blue) that were used to fit the atmospheric parameters (red). The yellow areas mark the line regions used for the fit, the grey areas show segments of 0.5\,\AA{} on each side of each line used for synthesising the spectra. In particular we show Fe\,I 6065, Fe\,II 6084, Ni\,I 6086, Fe\,I 6344, Si\,II 6347, Ca\,I 6472 and Si\,I 6527\,\AA{}.}
	\label{fit-w}
\end{figure*}

To create the synthetic spectra, we used the radiative transfer code \textit{SPECTRUM} with the \textit{ATLAS9.Kurucz} model atmosphere. Solar abundances were taken from \textit{Kurucz} and as line list we chose \textit{GESv5\_atom\_hfs\_iso.420\_92}.
For the fitting, we chose the red part of the spectrum ($6036-7003$\,\AA{}), because it yields a higher $S/N$. We chose a large number (>\,100) of temperature-sensitive Fe\,I, Fe\,II, Ca\,I, Ca\,II, Ni\,I, Si\,I and Si\,II lines as well as the wings of the H$\alpha$ absorption line, which has a big effect on the results because it spans a large wavelength range. After the automatic selection of the lines it had to be checked if they were suitable for fitting, i.e. they should reach back to the continuum within the segments that were drawn around each line. In some cases the segments were modified in order to reach that. In Fig.~\ref{fit-w}\hspace{-1.7mm}, we show examples of selected lines for TYC8152-550-1. The synthetic spectra are created in the segments (grey) while the differences to the observed spectrum are only computed within the line regions (yellow). If no continuum regions were available around a line, or if strong deviations between spectrum and fit were recognised within the line mask, the corresponding line was rejected.

As initial parameters for the fits, we used solar parameters as given in \citet{Blanco-Cuaresma2019} and the $T_{\text{eff}}$ of each star from \textit{Gaia} DR2. In each run, consisting of up to twelve iterations, it could be decided about which parameters should be fixed or free. As all parameters were unknown, we left all of them free, except either macroturbulence or $v_{\text{rot}}\sin(i)$, which are strongly correlated with each other. By running the fit several times with different combinations, we guaranteed that the fixed parameter among $v_t^{\text{mac}}$ and $v_{\text{rot}}\sin(i)$ was set to a reasonable value, while the other was left free. The limb darkening coefficient was fixed to 0.6, because changing it did not show any effect.

The resulting atmospheric parameters of the six Li-rich stars are displayed in Table~\ref{params}\hspace{-1.7mm}. The given intrinsic errors are computed from the covariance matrix which connects the errors of the individual parameters after the least-squares fitting \citep{Blanco-Cuaresma2014}. The flux errors for our spectra were calculated from gain and read noise of UVES with the given setup. For comparison, we also show $T_{\text{eff}}$ from \textit{Gaia} DR2. We give the results for both components of TYC159-251-1, where the fits were done individually after disentangling the spectra. Note that the disentangled spectra have lower $S/N$. The resulting $T_{\text{eff}}$ for TYC159-251-1\,B corresponds to SpT F9, but is also consistent with G0, which was used to disentangle the original spectrum. The derived $T_{\text{eff}}$ for TYC159-251-1\,A is fully consistent with the assumption of an F6 template spectrum. $v_t^{\text{mac}}=0\,$km\,s$^{-1}$ and $v_{\text{rot}}\sin(i)=0\,$km\,s$^{-1}$ were obtained for both components in all iterations, even when one or both of these parameters were left free.

\begin{center}
\begin{table*}%
\caption{The atmospheric parameters of the six Li-rich stars determined with \textit{iSpec}. The effective temperatures from \textit{Gaia} DR2 are also shown for comparison. A missing error indicates that the parameter was kept constant for all iterations. The results for TYC159-251-1 are given for both components individually.}
\label{params}
\centering
\begin{tabular*}{500pt}{cccccccc}
\hline
Name & $T_{\text{eff},Gaia}$ & $T_{\text{eff}}$ & $\log(g/\text{cm\,s}^{-2})$ & [M/H] & $v_t^{\text{mic}}$ & $v_t^{\text{mac}}$ & $v_{\text{rot}}\sin(i)$ \\
 & [K] & [K] & & & [km\,s$^{-1}$] & [km\,s$^{-1}$] & [km\,s$^{-1}$] \\
\hline
TYC159-251-1\,A & \hspace{1mm} $6257\pm391^*$ & $6308\pm59$ & $4.16\pm0.22$ & $-0.19\pm0.06$ & $1.49\pm0.24$ & 0 & 0 \\
TYC159-251-1\,B & \hspace{1mm} $6257\pm391^*$ & $6032\pm73$ & $3.94\pm0.27$ & $-0.16\pm0.07$ & $1.86\pm0.27$ & 0 & 0 \\
TYC159-343-1   & $7079\pm232$ & $6749\pm69$ & $2.77\pm0.35$ & $-0.57\pm0.08$ & $4.9\pm0.8$ & $16.4\pm2.0$ & 1.5 \\
TYC8150-2802-1 & $5853\pm100$ & $5857\pm75$ & $3.99\pm0.20$ & $-0.08\pm0.06$ & $1.28\pm0.19$ & 0 & 0 \\
TYC8150-3105-1 & $5614\pm228$ & $5956\pm73$ & $4.78\pm0.21$ & $-0.19\pm0.09$ & $1.6\pm0.5$ & $4.8$ & $13.0\pm0.9$ \\
TYC8152-1456-1 & $5806\pm77$ \hspace{1mm} & $5955\pm73$ & $4.14\pm0.23$ & $-0.09\pm0.07$ & $1.26\pm0.26$ & $0.8\pm1.9$ & 0 \\
TYC8152-550-1  & $6061\pm192$ & $6165\pm63$ & $4.26\pm0.24$ & $-0.27\pm0.06$ & $1.50\pm0.29$ & 0 & 0 \\
\hline
\end{tabular*}
\begin{tablenotes}
\item $^*$ \textit{Gaia} DR2 value for the unresolved system
\end{tablenotes}
\end{table*}
\end{center}

\subsection{Lithium equivalent widths and abundances}\label{LiEW}

The Li equivalent widths $EW_{\text{Li}}$ of the six Li-rich stars were measured with \textit{IRAF} by integrating over the area of the Li 6708\,\AA{} line with respect to the local continuum. The errors were calculated by adding the uncertainty in fitting the continuum and the error related to read-noise ($2.8\,\text{e}^-$) and gain ($0.52\,\text{e}^-/\text{ADU}$) of the instrument. The continuum error was derived by varying the flux level of the local continuum for the integration.

If we do not disentangle the spectrum of a spectroscopic binary, the equivalent widths will be underestimated due to the additional continuum flux from the other component. Therefore, we measured the $EW_{\text{Li}}$ for both components of TYC159-251-1 individually after disentangling. The absorption line is even stronger for the secondary component. As the F6 and G0 template spectra also show Li absorption, we removed the lines manually with \textit{IRAF} before subtracting the templates from the original spectrum. So it was guaranteed that the Li lines of both components remained in the disentangled spectra, where we could then measure the corresponding $EW_{\text{Li}}$.

The $EW_{\text{Li}}$ were converted to abundances by using the curves of growth given by \citet[Table 2, hereafter So93]{Soderblom1993}. The abundance scale is based on $\log(N_{\text{H}}) = 12$. The curves of growth are calculated for $T_{\text{eff}} =$~4000~$-$~6500\,K and local thermodynamic equilibrium. We estimated the Li abundances $\log(N_{\text{Li}})$ by choosing the values from So93 that best represent the given temperature and logarithmic equivalent width (for $EW_{\text{Li}}$ in m\AA{}) of the Li-rich stars, taking into account the errors. With $T_{\text{eff}}=6749\pm69$\,K, TYC159-343-1 is not covered by the $T_{\text{eff}}$ range in So93. Therefore, here we extrapolated the given abundances with a quadratic function and determined the Li abundance for $T_{\text{eff}} = 6750$\,K and $\log(EW_{\text{Li}})=1.50$, $\log(EW_{\text{Li}})=1.55$ and $\log(EW_{\text{Li}})=1.60$. The resulting abundance error, $\Delta \log(N_{\text{Li}})=^{+0.07}_{-0.08}$, contains the error from varying $EW_{\text{Li}}$ ($\Delta_{\text{var}}=^{+0.045}_{-0.048}$) and the root mean square (RMS) of the fit ($\Delta_{\text{RMS}}=0.029$ in all three cases). For the other Li-rich stars, the errors only come from the variation of $EW_{\text{Li}}$ and/or $T_{\text{eff}}$, according to the errors. The Li equivalent widths and abundances, together with temperatures, spectral types and magnitudes are listed in Table~\ref{Li}\hspace{-1.7mm}. The abundances can be compared to typical initial values, which are expected to be $\log(N_{\text{Li}})=3.1-3.3$ for population I stars \citep{Sestito2005}.

From \citet[Table 7]{Dantona1984} it can be seen that there is no significant Li depletion in stars with $M \gtrsim 1.2\,M_{\odot}$ if no additional mixing mechanisms occur. This corresponds to spectral type F7 (PM13). Therefore, the Li signal detected in TYC159-343-1 might not be conclusive to estimate its age and also TYC159-251-1 and TYC8152-550-1 have to be taken with care. Nevertheless, we will make an attempt to obtain rough age estimates for the Li-rich stars in the next subsection.

\begin{table*}
\caption{Temperatures, spectral types and magnitudes of the six Li-rich stars and the remaining observed early-type runaway star candidates. Effective Temperatures $T_{\text{eff}}$ of the Li stars and the stars in HB9 are from our parameter fits and from \citet{Bai2019}, respectively. The spectral types of the stars in Monoceros and Vela Jr. were taken from the \textit{Skiff} catalogue (Vo85, \citealt{Houk1978}). Conversion between $T_{\text{eff}}$ and SpT was done with PM13. \textit{G, BP} and \textit{RP} magnitudes are from \textit{Gaia} DR2, absolute $G$ magnitudes $M_G$ were calculated from \textit{Gaia G}, extinction and parallax and $G$ band extinctions are from the \textit{StarHorse} catalogue \citep{Anders2019}. The last two columns show Li equivalent widths $EW_{\text{Li}}$ and abundances $\log(N_{\text{Li}})$ for the six Li-rich stars. The abundance errors correspond to the minimum and maximum possible abundance from So93, taking into account the uncertainties in $EW_{\text{Li}}$ and $T_{\text{eff}}$. For TYC159-251-1, we give the values for the unresolved system as well as for the individual components, assuming that TYC159-251-1\,A contributes 66.1\,\% of the flux and TYC159-251-1\,B 33.9\,\%.}
\label{Li}
\centering
\begin{tabular*}{500pt}{cccccccccc}
\hline
Name & $T_{\text{eff}}$ & SpT & $G$ & $BP$ & $RP$ & $M_G$ & $A_G$ & $EW_{\text{Li}}$ & $\log(N_{\text{Li}})$ \\
 & [K] & & [mag] & [mag] & [mag] & [mag] & [mag] & [m\AA{}] & \\
\hline
Li-rich stars & & & & & & & & & \\
TYC159-251-1   & $6351\pm78$ &    F5--7    & 11.51 & 11.78 & 11.09     & $+2.61_{-0.19}^{+0.20}$ & $0.14_{-0.18}^{+0.20}$ & $17\pm2$ & $2.15^{+0.30}_{-0.05}$ \\
TYC159-251-1\,A   & $6308\pm59$ & F6--7    & 11.96 & 12.23 & 11.54  & $+3.06_{-0.19}^{+0.20}$ & $0.14_{-0.18}^{+0.20}$ & $22\pm3$ & $2.31^{+0.06}_{-0.11}$ \\
TYC159-251-1\,B   & $6032\pm73$ & F9--G0   & 12.68 & 12.96 & 12.26 & $+3.78_{-0.19}^{+0.20}$ & $0.14_{-0.18}^{+0.20}$ & $97\pm7$ &  $2.88^{+0}_{-0.08}$ \\
TYC159-343-1   & $6749\pm69$ &    F2--4    & 10.87 & 11.08 & 10.54     & $+1.24_{-0.23}^{+0.29}$ & $0.33_{-0.22}^{+0.28}$ & $35\pm5$ & $2.91^{+0.07}_{-0.08}$ \\
TYC8150-2802-1 & $5857\pm75$ &    G0--G2   & 12.03 & 12.36 & 11.55    & $+4.53_{-0.13}^{+0.16}$ & $0.04_{-0.13}^{+0.15}$ & $50\pm3$ & $2.25^{+0.03}_{-0.06}$ \\
TYC8150-3105-1 & $5956\pm73$ &    F9.5--G1 & 11.94  & 12.26 & 11.42 & $+4.34_{-0.16}^{+0.11}$ & $0.22_{-0.16}^{+0.11}$ & $177\pm7$ \hspace{1mm} & $3.37\pm0$ \\
TYC8152-1456-1 & $5955\pm73$ &    F9.5--G1 & 12.47 & 12.81 & 11.97  & $+3.35_{-0.26}^{+0.21}$ & $0.22_{-0.25}^{+0.21}$ & $66\pm4$ & $2.59^{+0.07}_{-0}$ \\
TYC8152-550-1  & $6165\pm63$ &    F7--F9   & 11.92 & 12.21 & 11.48    & $+3.52_{-0.17}^{+0.19}$ & $0.33_{-0.17}^{+0.19}$ & $33\pm4$ & $2.47^{+0.05}_{-0.26}$ \\
\hline
HB9 & & & & & & & & & \\
TYC3344-235-1 & $7707\pm377$ & A5--F0 & 11.07 & 11.38 & 10.61 & $-0.60_{-0.57}^{+0.21}$ & $1.45_{-0.57}^{+0.19}$ & -- & -- \\
TYC3344-679-1 & $7591\pm435$ & A6--F0 & 11.40 & 11.72 & 10.92 & $-0.28_{-0.51}^{+0.16}$ & $1.69_{-0.49}^{+0.09}$ & -- & -- \\
TYC3344-683-1 & $7490\pm360$ & A7--F0 & 12.33 & 12.58 & 11.94 & $+2.40_{-0.24}^{+0.30}$  & $0.35_{-0.23}^{+0.29}$ & -- & -- \\
TYC3344-553-1 & $7588\pm249$ & A7--A9 & 10.87 & 11.08 & 10.53 & $+1.06_{-0.24}^{+0.46}$  & $0.73_{-0.24}^{+0.45}$ & -- & -- \\
\hline
Monoceros Loop & & & & & & & & & \\
TYC159-2771-1 & $10700_{-1250}^{+2550}$ & B8--A0 & 12.01 & 12.13 & 11.78 & $+1.11_{-0.25}^{+0.23}$ & $0.29_{-0.22}^{+0.20}$ & -- & -- \\
HD\,261359 & $10700_{-1250}^{+2550}$    & B8--A0    & 11.79 & 11.87 & 11.62 & $+0.89_{-0.26}^{+0.26}$ & $-0.04_{-0.20}^{+0.20}$ & -- & -- \\
HD\,261393 & $15700_{-1450}^{+1150}$    & B4--B6    & 10.05 & 10.07 & 10.02 & $-0.61_{-0.24}^{+0.17}$ & $0.16_{-0.20}^{+0.12}$ & -- & -- \\
TYC159-2337-1 & $10700_{-1250}^{+2550}$ & B8--A0 & 11.80 & 11.88 & 11.61 & $-0.04_{-0.27}^{+0.27}$ & $0.27_{-0.13}^{+0.12}$ & -- & -- \\
TYC159-2671-1 & $12500_{-1950}^{+1750}$ & B7--B9 & 12.25 & 12.31 & 12.08 & $+1.25_{-0.22}^{+0.19}$ & $0.18_{-0.19}^{+0.15}$ & -- & -- \\
\hline
Vela Jr. & & & & & & & & & \\
HD\,76060 & $12500_{-950}^{+1750}$ & B7--B9 & 7.85 & 7.83 & 7.92 & $-0.63_{-0.11}^{+0.11}$ & $0.05_{-0.10}^{+0.10}$ & -- & -- \\
\hline
\end{tabular*}
\end{table*}

\subsection{Age estimation}\label{ages}

\begin{figure*}
	\centerline{\includegraphics[width=180mm]{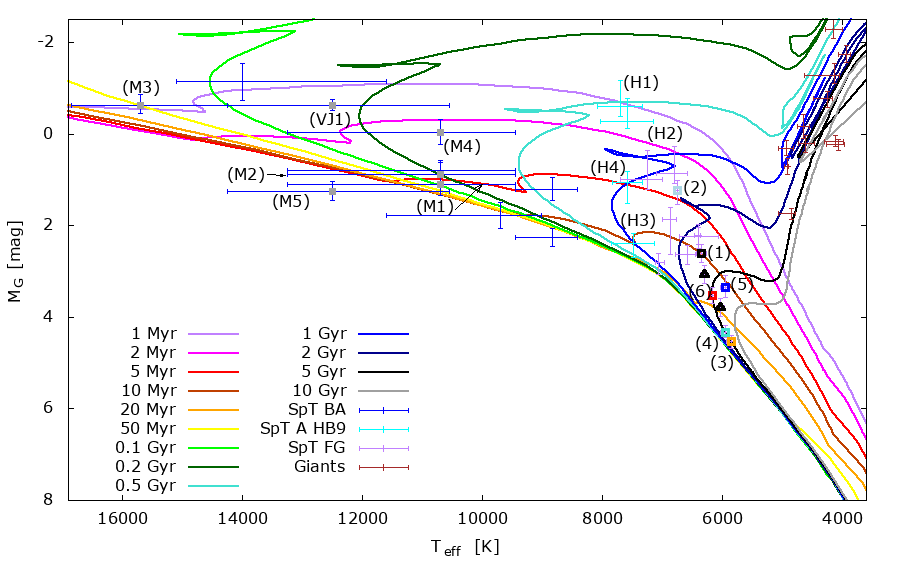}}
	\caption{HRD with absolute \textit{G}-band magnitudes of stars observed with UVES/VLT and HDS/Subaru. For an age estimate, we show PARSEC isochrones \citep{Bressan2012} with metallicity $Z=0.152$, corresponding to solar metallicity. For the early-type stars (blue), we used spectral types from the \textit{Skiff} catalogue (\citealt{Skiff2009} and references therein) and corresponding $T_{\text{eff}}$ ranges from PM13. The remaining candidates in the Monoceros Loop (M) and Vela Jr. (VJ) are marked by grey squares and labeled as (M1) TYC159-2771-1, (M2) HD\,261359, (M3) HD\,261393, (M4) TYC159-2337-1, (M5) TYC159-2671-1, (VJ1) HD\,76060. For stars located in HB9 (cyan), labeled as (H1) TYC3344-235-1, (H2) TYC3344-679-1, (H3) TYC3344-683-1 and (H4) TYC3344-553-1, we checked the $T_{\text{eff}}$ from \citet{Bai2019}, which are used here for the stars with spectral type late-A. For stars with $T_{\text{eff}}\lesssim7500$\,K (SpT FG) and no Li (purple), we use $T_{\text{eff}}$ from \textit{Gaia} DR2. For the Li-rich stars, marked with coloured squares, we show the $T_{\text{eff}}$ from our parameter fits. They are labeled as (1) TYC159-251-1, (2) TYC159-343-1, (3) TYC8150-2802-1, (4) TYC8150-3105-1, (5) TYC8152-1456-1 and (6) TYC8152-550-1. TYC159-251-1 is a binary and the HRD positions of the individual components correspond to the black triangles.}
	\label{HRD}
\end{figure*}

\begin{figure*}
	\centerline{\includegraphics[width=180mm]{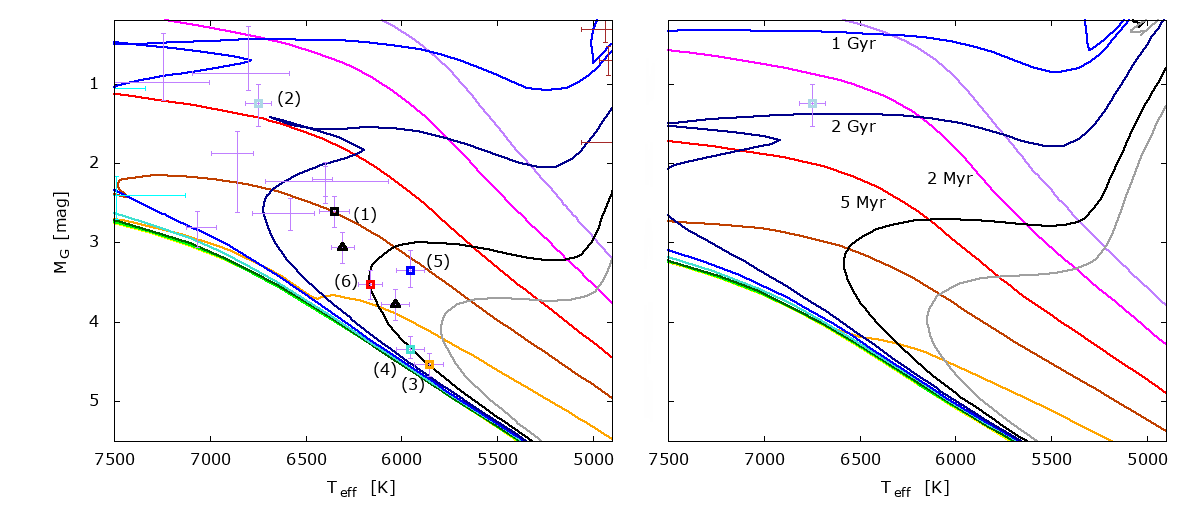}}
	\caption{Left: Zoom-in of Fig.~\ref{HRD}\hspace{-1.7mm}, centered on the Li-rich stars. Right: HRD with absolute \textit{G}-band magnitudes of TYC159-343-1. For an age estimate, we show PARSEC isochrones \citep{Bressan2012} with metallicity [M/H]\,$=-0.57$. This value and the $T_{\text{eff}}$ were determined from our parameter fits. The axis scales are similar in both panels. For the key, see Fig.~\ref{HRD}\hspace{-1.7mm}.}
	\label{HRD_zoom}
\end{figure*}

\begin{figure}
	\centerline{\includegraphics[width=88mm
	]{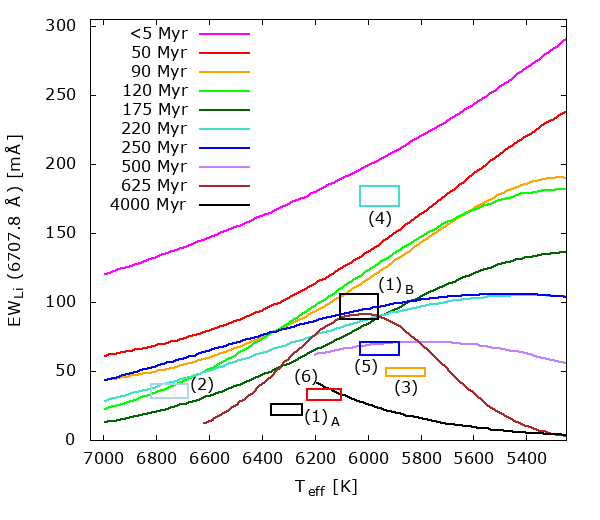}}
	\caption{$T_{\text{eff}}$ versus $EW_{\text{Li}}$ for the six Li-rich stars, shown with coloured error boxes and labeled as (1)$_\text{A}$ TYC159-251-1\,A, (1)$_\text{B}$ TYC159-251-1\,B, (2) TYC159-343-1, (3) TYC8150-2802-1, (4) TYC8150-3105-1, (5) TYC8152-1456-1 and (6) TYC8152-550-1. For TYC159-251-1, we only show the individual components. The curves represent open clusters with different ages and were fitted by Eric Mamajek to the data of the individual cluster stars. The original plot can be found under \href{http://www.pas.rochester.edu/~emamajek/images/li.jpg}{http://www.pas.rochester.edu/$\sim$emamajek/images/li.jpg}.}
	\label{Mamajek}
\end{figure}

In the following, we will compare two methods to estimate the ages of our targets. Firstly, we will use the positions of the stars in the HRD, where we can compare them to isochrones (Figs.~\ref{HRD}\hspace{-1.7mm} and \ref{HRD_zoom}\hspace{-1.7mm}). Then we perform the Li test by using the determined $T_{\text{eff}}$ and $EW_{\text{Li}}$ (or $\log(N_{\text{Li}})$) to compare them to stars of clusters with known ages (Figs.~\ref{Mamajek}\hspace{-1.7mm} and \ref{Li_clusters}\hspace{-1.7mm}).

\subsubsection{HRD isochrone placement}

In the HRD (Fig.~\ref{HRD}\hspace{-1.7mm}, for a zoom-in of the Li-rich stars see Fig.~\ref{HRD_zoom}\hspace{-1.7mm}), we show the stars observed with UVES/VLT and HDS/Subaru, using $T_{\text{eff}}$ either from \textit{Gaia} DR2, from their spectral type as given in the \textit{Skiff} catalogue \citep{Skiff2009}, from \citet{Bai2019} or from our parameter fits. Using \textit{Gaia} DR2 parallaxes and \textit{StarHorse} extinctions \citep{Anders2019}, we converted the apparent magnitudes $G$ to absolute magnitudes $M_G$. For the $G$-band extinctions, no errors are given in the \textit{StarHorse} catalogue. We converted the $V$ band extinction errors to $G$ with the relation $\Delta A_G=0.77\Delta A_V$ \citep{Mugrauer2019}. $\Delta M_G$ was then determined with Gaussian error propagation.

We compare our stars to isochrones which were calculated from PARSEC models \citep{Bressan2012}, reflecting the whole stellar evolution until the first thermal pulse. We used solar metallicity, $Z=0.152$, which corresponds to $[\text{M/H}]=0.015$. This is close to $[\text{M/H}]=-0.20\pm0.13$, the average metallicity of our targets as given in the \textit{StarHorse} catalogue. The corresponding shift of isochrone positions is rather small compared to the temperature and magnitude errors for early- and late-type stars, respectively. However, TYC159-343-1 has a significantly lower metallicity of $[\text{M/H}]=-0.57\pm0.08$. Therefore, for this target we plotted seperate isochrones to determine its age more precisely (see Fig.~\ref{HRD_zoom}\hspace{-1.7mm} and Table~\ref{HRD_ages}\hspace{-1.7mm}).

We directly see that the red giants are located on the giant branch of the isochrones in Fig.~\ref{HRD}\hspace{-1.7mm}, consistent with ages between 0.2\,Gyr and $>$\,10\,Gyr. The Li-rich stars are mostly located relatively close to the ZAMS, only TYC159-343-1 is located close to the isochrone for 5\,Myr due to its high brightness.

TYC159-251-1 was discovered by us to be a SB2. Photometrically unresolved binarity changes the positions in the HRD. The brightness is then overestimated, so the star is located significantly above the ZAMS. We correct for this effect by assuming that the primary component contributes 66.1\,\% and the secondary 33.9\,\% of the flux. This adds 0.45\,mag and 1.17\,mag, respectively, to their $M_G$ magnitude, shifting them back towards the ZAMS. Furthermore, the revised atmospheric parameter fits for the individual components yield lower effective temperatures. The two individual components differ by 276\,K and 0.72\,mag.

The positions of the early-type stars in the HRD have to be taken with care because \textit{Gaia} temperature estimates are trained only up to $T_{\text{eff}}=10000$\,K and tend to be underestimated when approaching this limit. Therefore, for the early-type stars observed with UVES, we adopted the spectral types from the \textit{Skiff} catalogue (\citealt{Skiff2009}). Only one reference is given for each star (\citealt{Houk1978} for HD76060, \citealt{Voroshilov1985}, hereafter Vo85, for the ten other early-type stars) and no error ranges are given there, so we assume them to be $\pm1$ subclass. The corresponding temperatures were inferred from PM13. Although these stars were also observed by us, the spectra were not suited for determining the spectral type. Our spectral range of $5000-7000$\,\AA{} is quite narrow and misses most spectral lines that are usually taken for the analysis of early-type stars. Additionally, the few lines we found are often affected by strong rotational broadening. Nevertheless, we checked the \textit{Skiff} spectral types for consistency by comparing our spectra to standard star spectra observed with the \textit{Fibre Linked ECHelle Astronomical Spectrograph} (FLECHAS, \citealt{Mugrauer2014}), operated at the 0.9\,m telescope of the University Observatory Jena. We mainly used the He\,I line at 5015.7\,\AA{} (visible up to B9) and the Fe lines at $5018.0-5018.4$\,\AA{} (visible from A0 on). Our results are largely consistent with the spectral types given by Vo85, with deviations of up to four subclasses.

For the stars observed with HDS/Subaru, from comparison with the spectra we found that the $T_{\text{eff}}$ in \textit{Gaia} DR2 are underestimated. For these targets, which are not listed in the \textit{Skiff} catalogue, instead we used the values given in \citet{Bai2019}, which fit best to our spectra. Therefore, they probably have spectral types A5\,$-$\,F0. However, their absolute magnitudes differ a lot from each other, indicating different ages.

HD\,261393, suggested by Bo17 as the best candidate in the Monoceros Loop, has $T_{\text{eff}}=14250-16850$\,K (Vo85) and $M_G\approx-0.6$\,mag. Therefore, it is probably an evolved but still young star of $\lesssim$\,105\,Myr, consistent with being the ejected companion of a SN-progenitor ($t\lesssim32$\,Myr).

\begin{table}
\begin{threeparttable}
\caption{Estimated HRD ages (see Figs.~\ref{HRD}\hspace{-1.7mm} and \ref{HRD_zoom}\hspace{-1.7mm}) for the six Li-rich stars, including both components of the SB2 TYC159-251-1, as well as for the remaining observed runaway star candidates in HB9, Monoceros Loop and Vela Jr. If a star is consistent with the zero-age main-sequence, only a lower limit is given in column 2. Otherwise, the values in column 2 represent the case that the star is on its pre-MS and the values in column 3 represent the case that the star is on its terminal-age- or post-MS.}
\label{HRD_ages}
\centering
\begin{tabular*}{220pt}{ccc}
\hline
Name & $t_{\text{HRD}}$ [Myr] & $t_{\text{HRD}}$ [Myr] \\
 & if (pre-)MS & if post-MS \\
\hline
Li-rich stars & & \\
TYC159-251-1\,A & 11 -- 16 & 3200 -- 4500 \\
TYC159-251-1\,B & 14 -- 30 & 4800 -- 7500 \\
TYC159-343-1*  & 2.1 -- 3.8 & 1600 -- 2300 \\
TYC8150-2802-1 & $>30$ & -- \\
TYC8150-3105-1 & $>25$ & -- \\
TYC8152-1456-1 & \hspace{1mm} 8 -- 16 & 5000 -- 8000 \\
TYC8152-550-1  & 13 -- 20 & 3800 -- 6000 \\
\hline
HB9 & & \\
TYC3344-235-1 & $<1.6$ & 380 -- 650 \\
TYC3344-679-1 & $2.0$ & 490 -- 800 \\
TYC3344-683-1 & $>9$ & -- \\
TYC3344-553-1 & 3.7 -- 7 \hspace{1mm} & \hspace{1mm} 850 -- 1300 \\
\hline
Monoceros Loop & & \\
TYC159-2771-1 & $>4.2$ & -- \\
HD\,261359 & $>3.5$ & -- \\
HD\,261393 & $<110$ & -- \\
TYC159-2337-1 & 1.6 -- 3.3 & \hspace{1mm} 50 -- 490 \\
TYC159-2671-1 & $>5$ & -- \\
\hline
Vela Jr. & & \\
HD\,76060 & 1.1 -- 1.8 & \hspace{1mm} 90 -- 350 \\
\hline
\end{tabular*}
\begin{tablenotes}
\item * HRD ages determined with isochrones for $[\text{M/H}]=-0.57$.
\end{tablenotes}
\end{threeparttable}
\end{table}

\subsubsection{Lithium test}

Fig.~\ref{Mamajek}\hspace{-1.7mm} shows $T_{\text{eff}}$ and $EW_{\text{Li}}$ for stars in several open clusters of different ages\footnote{\href{http://www.pas.rochester.edu/~emamajek/images/li.jpg}{http://www.pas.rochester.edu/$\sim$emamajek/images/li.jpg}}. Eric Mamajek fitted polynomials to the data of cluster stars, allowing a rough localisation of ages in the $T_{\text{eff}}-EW_{\text{Li}}$ space. By comparing the values of the Li-rich stars, with $T_{\text{eff}}$ from our parameter fits, to these curves, we obtained the age ranges displayed in Table~\ref{Mamajek_ages}\hspace{-1.7mm}. Further uncertainty comes from the unknown initial Li abundances of our targets, depending on the metallicity \citep{Lambert2004}. If the initial abundance was lower than in the case of the cluster stars, less time would have been necessary to reach the current abundance. Also, the data that were used to fit the polynomials show a large scatter, e.g. due to different metallicities and rotational velocities, as well as possible systematic effects like unresolved binarity or starspots. Therefore, quantifying ages for individual stars is unreliable and the values given in Table~\ref{Mamajek_ages}\hspace{-1.7mm} should only be seen as a rough estimate.

\begin{table}
\caption{Ages of the six Li-rich stars, including both components of the SB2 TYC159-251-1, estimated from Fig.~\ref{Mamajek}\hspace{-1.7mm}.}
\label{Mamajek_ages}
\centering
\begin{tabular*}{140pt}{cc}
\hline
Name  & $t_{\text{Li}}$ [Myr]\\
\hline
TYC159-251-1\,A & $>625$ \\
TYC159-251-1\,B & 90 -- 4000 \\
TYC159-343-1  & 90 -- 625 \\
TYC8150-2802-1 & 625 -- 4000 \\
TYC8150-3105-1 & 5 -- 50 \\
TYC8152-1456-1 & 250 -- 4000 \\
TYC8152-550-1  & $>625$ \\
\hline
\end{tabular*}
\end{table}

\subsubsection{Comparison}

We summarise the ages inferred from the two diagrams in Tables~\ref{HRD_ages}\hspace{-1.7mm} and \ref{Mamajek_ages}\hspace{-1.7mm}. TYC8150-2802-1 and TYC8150-3105-1 are consistent with the ZAMS in Fig.~\ref{HRD}\hspace{-1.7mm}. For TYC8150-2802-1, the low Li content indicates that it is an evolved star, at least a few hundred Myr old. TYC8150-3105-1, however, shows very strong Li absorption indicating that it is young. From combining both results, we obtain an age of $25-50$\,Myr, so we suggest that it is on its early MS. In the SB2 system TYC159-251-1, the primary is slightly above the ZAMS while the secondary is consistent with it. Still, their ages are consistent with each other. A young age of $\sim$15\,Myr can not be excluded from the HRD, but we consider it more likely that the binary is an evolved system having $t_{\text{HRD}}\approx4.5-4.8$\,Gyr because their Li content is too small for a young age.

TYC159-343-1 is a difficult case: From the parameter fits we obtained a very low metallicity of [M/H]\,$=-0.57\pm0.08$, so we inferred its HRD-age from comparison to low-metallicity isochrones (see Fig.~\ref{HRD_zoom}\hspace{-1.7mm}). Its high brightness of $M_G=1.24\pm0.19$\,mag indicates an age of only $2.1-3.8$\,Myr, if it is pre-MS. However, it is more likely an evolved star with an age of $1.6-2.3$\,Gyr. The Li abundance, $\log(N_{\text{Li}})=2.91\pm0.05$, is not a good age tracer in this case, because of two caveats: (i) The low metallicity also means that the star probably had a low initial Li abundance, e.g. $\log(N_{\text{Li}})=2.64\pm0.07$ according to \citet[Table 2]{Lambert2004}, which is even below the measured abundance. A lower initial abundance reduces the time which is necessary to reach the current abundance and therefore the age. (ii) TYC159-343-1 has spectral type F2\,$-$\,4, meaning that no significant Li depletion is expected from standard mixing mechanisms \citep{Dantona1984}. In Fig.~\ref{Mamajek}\hspace{-1.7mm}, the star lies significantly below the 5\,Myr curve and the age can be given as $t_{\text{Li}}=90-625$\,Myr. However, due to the caveats stated above, we rather rely on the age ranges estimated from Fig.~\ref{HRD_zoom}\hspace{-1.7mm}.

The HRD positions of TYC8152-1456-1 and TYC8152-550-1 also allow both possibilities, i.e. they could be either pre- or post-ZAMS. Applying the Li test to them excludes the possibility that they are very young. So, they are probably evolved stars, unrelated to the birth association of the Vela Jr. progenitor, whereas TYC8150-3105-1 could be from the same stellar group that gave birth to the Vela progenitor.

\subsubsection{The lithium depletion gap}

The Hyades, at an age of $\sim$625\,Myr, show an interesting feature that contradicts standard mixing mechanisms: Stars of $T_{\text{eff}}\approx6600\pm300$\,K show Li depletions which are enhanced by a factor $\sim$100 compared to neighbouring hotter and cooler stars \citep{Boesgaard1986, Thorburn1993, Deliyannis2019}. In a weaker extent, this feature can also be seen in younger clusters like M35 ($\sim$175\,Myr), where the gap is just forming \citep{Steinhauer2003, Steinhauer2004}. The gap of enhanced Li depletion is caused by non-standard mixing on the MS, acting predominantly in the above stated $T_{\text{eff}}$ regime. Slow mixing, e.g. by rotational and/or gravitational instabilities, and diffusion can play a role in the formation of the gap \citep{Steinhauer2004}. Recent works \citep{Steinhauer2021} show the effect of rotational mixing for the formation of the gap. The authors find that cluster stars with a high $v_{\text{rot}} \sin(i)$ have a more strongly decreasing Li abundance with age. Therefore, observations of stars within the gap and comparing their Li abundances and $v_{\text{rot}} \sin(i)$ to clusters of known ages can give a further estimate of their ages. Unfortunately, the method is not applicable for us due to the insignificant measurements of $v_{\text{rot}} \sin(i)$ and the unknown initial Li abundances. Additionally, the method only works for stars within the Li depletion gap and only for ages of $\sim$100~$-$~650\,Myr. However, by comparing the measured abundances to cluster data we can still obtain a consistency check of the ages given in Tables~\ref{HRD_ages}\hspace{-1.7mm} and \ref{Mamajek_ages}\hspace{-1.7mm}. This comparison also works for cooler stars, while it also suffers from the same caveats as described for the Li test above, e.g. the large scatter of the $\log(N_{\text{Li}})$.

\begin{figure}
	\centerline{\includegraphics[width=88mm]{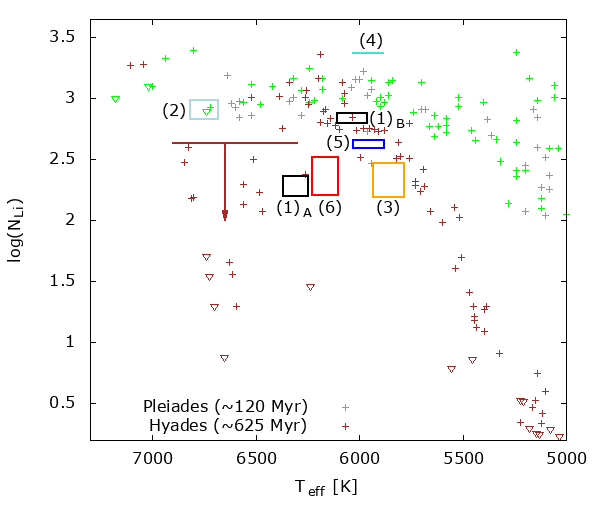}}
	\caption{Lithium abundances of Pleiades and Hyades stars. The data for the Pleiades (green) were taken from \citet[Table 1]{Soderblom1993}, including data from \citet{Boesgaard1988, Butler1987} and \citet{Pilachowski1987}. The data for the Hyades (brown) were taken from \citet{Thorburn1993} and \citet{Boesgaard1986}. Inverted triangles show upper limits in the case of non-detections. Our Li targets, shown with coloured squares representing the error boxes, are labeled as in Fig.~\ref{Mamajek}\hspace{-1.7mm}. The stars (1)$_\text{A}$ TYC159-251-1\,A and (2) TYC159-343-1 are consistent with the Li depletion gap, which is marked by the brown arrow and the attached horizontal bar.}
	\label{Li_clusters}
\end{figure}

In Fig.~\ref{Li_clusters}\hspace{-1.7mm} we compare our Li targets to data of the clusters Pleiades (\citealt{Butler1987}; \citealt{Pilachowski1987}; \citealt{Boesgaard1988}; So93) and Hyades \citep{Thorburn1993, Boesgaard1986}. The Li depletion gap of the Hyades is indicated by the brown arrow and the attached horizontal bar at $T_{\text{eff}}=6300-6900$\,K. Note that the hot edge of the gap is steeper than the cold edge, where a larger scatter is observed. The $T_{\text{eff}}$ of TYC159-251-1\,A and TYC159-343-1 are consistent with the Li gap. TYC159-251-1\,A lies at the cold edge of the gap and its Li abundance is lower than of Hyades stars with the same $T_{\text{eff}}$. Therefore, the Hyades age could be seen as a lower limit for the SB2 TYC159-251-1. TYC159-343-1 has one of the highest Li abundances among our sample, despite its low metallicity. Its Li abundance is comparable to Pleiades stars in this $T_{\text{eff}}$-region, so the Pleiades age ($\sim$125\,Myr) can be seen as an upper limit. Possibly the star is younger than inferred from Fig.~\ref{Mamajek}\hspace{-1.7mm} ($90-625$\,Myr), where we did not consider the metallicity.

The other four Li-rich targets have lower $T_{\text{eff}}$ and do not fall in the Li depletion gap. The abundances of TYC8150-2802-1, TYC8152-1456-1 and TYC8152-550-1 are consistent with the Hyades or older, whereas TYC8150-3105-1 has more Li than comparable Pleiades stars. Therefore, it is very young, consistent with what we found from Fig.~\ref{Mamajek}\hspace{-1.7mm}.

\section{Discussion}\label{sec5}

\subsection{Overview of the candidates}

\begin{center}
\begin{table*}
\begin{threeparttable}%
\caption{Numbers of runaway star candidates with $G<17$\,mag in the SNRs covered in this work. Columns 2 and 3 give the numbers of candidates identified in \textit{Gaia} DR1 TGAS and DR2, respectively. Column 4 gives the number of observed stars and if they were observed with UVES/VLT (U) or with HDS/Subaru (H). These observations are all based on the DR1 selection, whereupon \text{Gaia} DR2 was used for the further characterisation of the targets. Column 5 gives the names of associated neutron stars and column 6 the numbers of remaining candidates that could not be excluded during the analysis.}
\label{cands}
\centering
\begin{tabular*}{500pt}{ccccccc}
\hline
SNR Name & DR1 & DR2 & Obs. & PSR & Rem. cands. & Additional info   \\
\hline
G074.0$-$08.5 & 16 & 19 & 0 & -- & 19 & \\
G160.9+02.6 & 6 & 1 & 6 (H) & SGR 0501+4516(?) & 5 & SGR \& PSR prob. unrelated  \\
G180.0$-$01.7 & 3 & -- & 3 (U) & J0538+2817 & 1 & HD\,37424 confirmed as runaway$^{1}$ \\
G205.5+00.5 & 21 & 116$^{**}$ & 21 (U) & -- & 5 & incl. HD\,261393$^{2}$ \\
G260.4$-$03.4 & 2 & 9 & 2 (U) & -- & 9 & PSR J0821$-$4300 prob. unrelated\\
G263.9$-$03.3 & 2 & 1 & 2 (U) & J0835$-$4510 & 1 & \\
G266.2$-$01.2 & 5 & 12 & 5 (U) & -- & 13 & PSR J0855$-$4644 prob. unrelated  \\
G330.0+15.0 & 6 & 3 & 0 & -- & 3 & \\
G111.7$-$02.1$^*$ & 0 & 0 & 0 & CCO J232327.9+584842 & 0 &  \\
G130.7+03.1$^*$ & 0 & 0 & 0 & J0205+6449 & 0 & No cand. consistent with PSR \\
G184.6$-$05.8$^*$ & 0 & 0 & 0 & J0534+2200 & 0 &  \\
G347.3$-$00.5$^*$ & 2 & 18 & 0 & CCO(?) & 18 & \\
\hline
\end{tabular*}
\begin{tablenotes}
\item $^*$ Historical SNR; $^{**}$ not suggested for observations, more precise constraints required; $^1$ \citet{Dincel2015};  $^2$ \citet{Boubert2017}
\end{tablenotes}
\end{threeparttable}
\end{table*}
\end{center}

We list in Table~\ref{cands}\hspace{-1.7mm} the SNRs with their corresponding numbers of runaway star candidates from the different selection steps and if they were already observed, as well as additional information (e.g. the best candidates). A list of the individual candidate stars can be found in Table~\ref{Runawaylist}\hspace{-1.7mm}.

Note that for the candidates selected from \textit{Gaia} DR2 the allowed locations around the SNR centres at the time of the SN were reduced from an error estimate based on the determination of the Vela SNR centre to a smaller error based on the description in \citet{Green2009} (see Section~\ref{sec2.2}). This was necessary to limit the high number of stars from DR2 more strictly, in order to create feasible projects for follow-up observations. In both cases the errors scale with the size of the SNR. Trajectories of associated pulsars were considered for the DR2 selection.

In S147, we can confirm the B0.5\,V star HD\,37424 based on DR2 data. Its position at the time of the SN is well within the 1\,$\sigma$ error ellipse of the GC used for the DR1 selection and only slightly outside (by 8.5\,\%) of the more strict error ellipse used for the DR2 selection. We also checked for the possibility that more than one star might be ejected from a SN in a multiple system. But the three other runaway star candidates besides HD\,37424, which projected trajectories could originate from the GC and which were observed with UVES, are neither consistent with PSR J0538+2817 nor with HD\,37424.

Two of the DR2 candidates in the Lupus Loop are particularly promising. Their positions at the time of the SN were $\ang{;1.1;} \pm \ang{;6.3;}$ and $\ang{;1.7;}\pm \ang{;6.6;}$ off the GC, respectively (see Table~\ref{Runawaylist}\hspace{-1.7mm}) and both have very high proper motions clearly directing away from the centre. Spectroscopic follow-up observations are highly suggested for them.

The historical SNRs generally have very small diameters due to their low age. This reduces the number of stars in the cone search. Just as \citet{Fraser2019}, for Cas A and the Crab Nebula we did not find any candidates. For SNR G130.7+03.1 (from SN 1181) we found one star with $G<17.0$\,mag to be consistent with the GC, but it was ruled out due to the missing kinematical consistency with PSR J0205+6449. SNR G347.3$-$00.5 probably originates from SN 393 and is much larger than the other three historical SNRs ($\Theta = \ang{;65;} \times \ang{;55;}$). Here we found 18 candidates to be consistent with the GC. The other SNRs are described in more detail in the following subsection.\\~\\

\subsection{Individual SNRs}

\subsubsection{HB9}

HB9 is a $\ang{;140;} \times \ang{;120;}$ SNR on the northern hemisphere, where we found six runaway star candidates around the GC from \textit{Gaia} DR1, which were observed with HDS/Subaru. The adopted SNR distance and age are $d=0.8\pm0.4$\,kpc and $t=5.5\pm1.5$\,kyr, respectively \citep{Leahy2007}. Among the observed candidates, we found two giants. The parameters of the other four are given in Tables~\ref{veloc}\hspace{-1.7mm} and \ref{Li}\hspace{-1.7mm}. They have a large range of effective temperatures in the literature, where the \textit{Gaia} DR2 values are probably underestimated, as can be seen from comparison with the spectra. Unfortunately, the chosen spectral range was not convenient for spectral type determination of early-type stars. Therefore, we rely on the $T_{\text{eff}}$ given by \citet{Bai2019}, which correspond to spectral type late-A. The RVs were calculated from the positions of individual absorption lines (see Section~\ref{sec4RV}), where the uncertainties are large due to the low number of available lines. The most precise value is given for TYC3344-235-1, which also shows the most promising kinematics. Although the proper motion is moderate, its high radial velocity (pointing towards us) gives it the highest peculiar velocity among all observed early-type stars ($v_{\text{pec}}=36.0\pm2.9$\,km\,s$^{-1}$). Its high distance of $d=1103^{+48}_{-45}$\,pc would then require that the SNR lies at the upper edge of its distance range. TYC3344-235-1 and TYC3344-679-1 have high luminosities of $34\,L_{\odot}$ and $20\,L_{\odot}$, respectively. From their spectra it is clear that they are not giants. They could be very young A-type stars but it is more likely that they are evolved, with ages between $\sim$0.2\,Gyr and 2\,Gyr, especially as TYC3344-235-1 is even located above the isochrone for 1\,Myr in Fig.~\ref{HRD}\hspace{-1.7mm}.

There are two neutron stars in the vicinity of HB9. For PSR J0502+4654, its high characteristic age of $\tau_c=1.81$\,Myr \citep{Manchester2005} and its proper motion indicate that it is unrelated. The magnetar SGR 0501+4516 is located $\sim$$\ang{1.38;;}$ to the south of the centre. If the location of the GC is close to the explosion site, an association would mean that SGR 0501+4516 travels with more than two times the largest 2D velocity of any other known pulsar \citep{Hobbs2005}, which makes the association very unlikely.

From \textit{Gaia} DR2, we found one fainter, yet unobserved object consistent with the GC. As both neutron stars are probably unrelated, we suggest that this is the most probable candidate. It has $G=16.27$\,mag and was located $\ang{;1.7;}\pm \ang{;3.0;}$ from the GC at the time of the SN. Its spectral type is K5.5\,$-$\,K7 according to its \textit{Gaia} DR2 $T_{\text{eff}}=4172_{-67}^{+184}$\,K. From its position in Fig.~\ref{HRDL}\hspace{-1.7mm} ($L=0.199\pm0.021\,L_{\odot}$), it should have a young age of $\sim$15~$-$~50\,Myr, so it could indeed be the ejected companion of the HB9 progenitor.

\subsubsection{Monoceros Loop}

The Monoceros Loop is one of the largest ($\Theta=\ang{;220;}$) and probably the oldest SNR of our sample, although the age determination is very uncertain ($30-150$\,kyr, \citealt{Welsh2001}). From \citet{Ferrand2012}, we adopted $d=1.44\pm0.54$\,kpc, as recent works \citep{Yu2019, Zhao2018} place it at $d=0.941_{-0.094}^{+0.096}$\,kpc or $d=1.257_{-0.101}^{+0.092}$\,kpc and $d=1.98$\,kpc, respectively. The discrepancy connects to the question if the Monoceros Loop interacts with the Rosette Nebula (3C\,163), an HII region bordering the SNR at the south-western edge.

We observed 21 stars with UVES/VLT in the Monoceros Loop, where no associated neutron star is known. Eight of those were ruled out because they were found to be giants. Among the others, the most interesting cases are (i) HD\,261393, a B5\,V star according to Vo85, which was proposed as the most likely runaway companion by Bo17, (ii) TYC159-251-1, which was newly discovered by us to be a double-lined spectroscopic binary and (iii) TYC159-343-1, an F2--F4 star. TYC159-343-1 and both components of TYC159-251-1 show some Li absorption. However, the ages inferred in Section~\ref{ages} are higher than what we expect for a BES runaway star. Also, from the updated distance to the Monoceros Loop and from using proper motions and parallaxes from \textit{Gaia} DR2, TYC159-343-1 ($d=0.723_{-0.021}^{+0.023}$\,kpc) and TYC159-251-1 ($d=0.564_{-0.014}^{+0.013}$\,kpc) turn out to be foreground stars, even when the lower limit of the SNR distance given by \citet{Yu2019} is extended to $2\,\sigma$ ($0.941\,\text{kpc}-0.188\,\text{kpc}=0.753$\,kpc).

Bo17 did not have \textit{Gaia} DR2, so we can still ask the question why HD\,261393 gained a higher probability in their work than TYC159-343-1. There are two reasons: (1) It is/was located closer to the GC. Bo17 used a prior to describe the location of the centre as a 2D-Gaussian with a full width at half maximum (FWHM) of $\tilde{\Theta} = \text{max}(\ang{;5;}, 0.05\,\Theta)$. This corresponds to $\tilde{\Theta} = \ang{;11;}$ in the case of the Monoceros Loop. The angular distance between TYC159-343-1 and the Green centre at the time of the SN was $\ang{;21;}\pm \ang{;22;}$, almost twice the FWHM. This compares to $\ang{;10;}\pm \ang{;25;}$ for HD\,261393. Note that the errors also take into account the uncertainty of the GC. (2) It is more massive. Multiplicity studies (e.g. \citealt{Pinsonneault2006}) have shown that massive stars tend to form binaries with similar masses. Therefore, Bo17 assign higher runaway probabilities to more massive, early-type stars. However, in wide systems the companions can form more independently and the distribution of mass ratios $q$ becomes almost consistent with random pairings from the initial mass function (IMF) \citep{Moe2017}. \cite{Sana2011} and \cite{Sana2012} present relatively flat mass ratio distributions down to $q=0.2$. Unfortunately, the regime of the most extreme mass ratios, $q<0.1$, is usually not covered by multiplicity studies, because the extreme flux ratios as well as decreasing RV precision in the case of early-type stars make low-mass companions hard to detect. However, there is no reason to assume that the mass ratio distributions should change for $q<0.1$.

Among the other observed stars in the Monoceros Loop, there is TYC159-1896-1 which has spectral type F3\,$-$\,9 (according to its $T_{\text{eff}}$ as listed in \textit{Gaia} DR2) and which does not show Li, hence is too old, and nine stars of spectral types mid-B to mid-A. Five of those turn out to be spatially inconsistent with the GC after reevaluating them with DR2 data.

So in total, we have five good runaway star candidates in the Monoceros Loop, including HD\,261393 and four stars of spectral types B7\,$-$\,A1, which trajectories are displayed in Fig.~\ref{Mono}\hspace{-1.7mm} and which kinematic data are noted in Table~\ref{veloc}\hspace{-1.7mm}. Their RVs were determined from the positions of individual absorption lines (see Section~\ref{sec4RV}). For TYC159-2671-1, H$\alpha$ was the only visible absorption line, so here the error was adopted from the highest error among the other stars. Their spectral types were adopted from Vo85, where the given range comes from our error estimate of $\pm1$ subclass.

A search for further targets in the DR2 catalogue yields more than 100 candidates with $G<17$\,mag, due to the large uncertainties of distance and age of the SNR. We abstain from listing them all in Table~\ref{Runawaylist}\hspace{-1.7mm} because it is not realistic to observe them all in the near future. It appears more fruitful to first concentrate on the brighter candidates observed with UVES/VLT.

Considering the large discrepancy of possible distances and ages, the identification of a runaway star would be particularly important for the Monoceros Loop to constrain these parameters significantly.

\subsubsection{Puppis A}

The Vela region, located on the southern sky in the Vela constellation, is a very extended and complex region with several stellar clusters, star-forming regions and three SNRs.

The most distant SNR in this region is Puppis A. Here, two runaway star candidates were observed with UVES, namely TYC7669-1336-1 and TYC7669-1414-1, which could both originate from the GC in projection. According to their $T_{\text{eff}}$ from \textit{Gaia} DR2, their spectral types are F3\,$-$\,F6 and F1\,$-$\,F2, respectively. No Li was found in the spectra. After reevaluating them with the DR2 parallax, it was found that they are too close to be consistent with the distance given for Puppis A ($d=1.3\pm0.3$\,kpc, \citealt{Reynoso2017}).

A search in \textit{Gaia} DR2 yields nine candidates with $G<17$\,mag, where eight of them have proper motions pointing to north-west, similar to most other stars in this region of the sky. Only one candidate, a K1.5\,$-$\,K3 star with $G=13.2$\,mag, has a higher proper motion and a slightly different direction compared to the other stars, pointing to west-north-west. It might, therefore, be the best candidate. It was located $\ang{;2.1;}\pm \ang{;2.7;}$ from the GC at the time of the SN. With $T_{\text{eff}} = 5042_{-180}^{+115}$\,K and $L=4.64\pm0.21\,L_{\odot}$, from Fig.~\ref{HRDL}\hspace{-1.7mm} it could be consistent with a young age of $\sim$1\,$-$\,2\,Myr, although it is more likely an evolved star with $4.5-11$\,Gyr.

\subsubsection{Vela}

With $d=0.275\pm0.025$\,kpc, Vela is the closest SNR, therefore having a large diameter of $\Theta=\ang{;255;}$. Its age is only poorly constrained to $t=18\pm9$\,kyr \citep{Ferrand2012}. However, the characteristic age of the Vela pulsar, $\tau_c=11.3$\,kyr, indicates that the SNR age could be close to the lower limit.

Two stars were observed, TYC8150-2802-1 and TYC8150-3105-1, both showing Li in their spectra. Their spectral types are G0\,$-$\,G2 and F9.5\,$-$\,G1, respectively. From the Li content and the position in the HRD, we can conclude that TYC8150-3105-1 has a relatively young age of $25-50$\,Myr, while TYC8150-2802-1 is probably an evolved MS star. Anyway, they cannot be associated runaway stars, because their trajectories are not consistent with the past position of the Vela pulsar (J0835$-$4510). TYC8150-3105-1 could be a member of the young association to which also the Vela progenitor belonged, whereas TYC8150-2802-1 is probably an interloper.

\begin{figure}
\includegraphics[width=84mm]{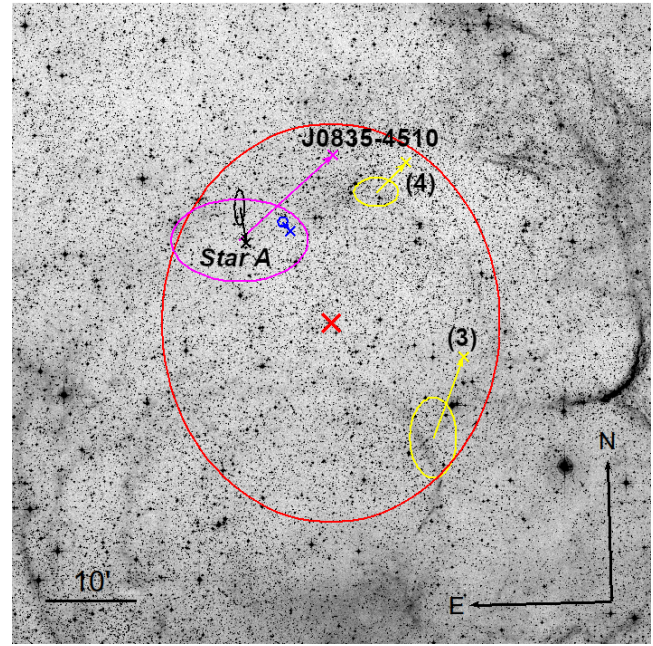}
\caption{The central $\ang{;70;} \times \ang{;70;}$ of the Vela SNR. The red labels mark the geometric centre and its error ellipse, the yellow labels the Li-rich stars observed in ESO P100, namely (3) TYC8150-2802-1 and (4) TYC8150-3105-1. The crosses mark the current positions, connected to the positions at the time of the SN (with error ellipse) by an arrow in proper motion direction. The motion of the Vela PSR (J0835$-$4510) is shown in magenta, our runaway candidate from \textit{Gaia} DR2 in blue and the candidate identified by FB19 in black. Background image from ESO DSS-2-red.}
\label{Velacands}
\end{figure}

Fig.~\ref{Velacands}\hspace{-1.7mm} shows the motion of the Li-rich stars in yellow and the Vela pulsar in magenta. Two further candidates are marked, found in \textit{Gaia} DR2, which are consistent with the Vela pulsar: The star marked in blue was identified by us and could be observed in the near future. The star marked in black was identified by \citet{Fraser2019} (hereafter FB19), therein denoted as \textit{Star A}. It is closer to the nominal past position of the Vela pulsar, but with $G=20.1$\,mag it is much too faint to obtain a spectrum with sufficient $S/N$. FB19 note that it is unlikely that the ejected companion of the Vela SN progenitor would be such faint (absolute magnitude $M_G=12.7$\,mag), so probably \textit{Star A} is an unrelated background star. FB19 used the distribution of runaway star velocities given by Rz19 to constrain their search radius, which limits their selection more strictly than the $v_{\text{max}}=1000\,\text{km\,s}^{-1}$ used by us. Therefore, the star marked in blue in Fig.~\ref{Velacands}\hspace{-1.7mm} was not found by them.

\subsubsection{Vela Junior}

Vela Jr. is located about $\ang{3;;}$ east of Vela. With an age between 2.4\,kyr and 5.1\,kyr it is younger than Vela and its distance is between 0.5\,kpc and 1.0\,kpc \citep{Allen2015}.

Five stars were observed with UVES: TYC8152-120-1 is a mid-F star, where no Li was discovered. TYC8152-104-1 is a K giant, and TYC8152-1456-1 (F9.5\,$-$\,G1) and TYC8152-550-1 (F7\,$-$\,F9) show Li absorption in their spectra. However, the measured abundances indicate that they are much too old to be runaway stars associated with the Vela Jr. progenitor (see Figs.~\ref{Mamajek}\hspace{-1.7mm}, \ref{Li_clusters}\hspace{-1.7mm}, Table~\ref{Mamajek_ages}\hspace{-1.7mm}). HD\,76060 is the brightest star of our observed sample with $G=7.85$\,mag. Its spectral type is B8\,IV--V, according to \citet{Houk1978}. From its \textit{Gaia} DR2 parallax, we infer a distance of $d=0.486_{-0.010}^{+0.011}$\,kpc, which could still be consistent with the lower distance limit of Vela Jr. Its position at the time of the SN is consistent with the GC according to Eqn.~\ref{SNRerror} and its peculiar velocity of $31.2\pm2.7$\,km\,s$^{-1}$ (see Table~\ref{veloc}\hspace{-1.7mm} for further parameters) is relatively high. Its age, according to Fig.~\ref{HRD}\hspace{-1.7mm}, is either $1.1-1.8$\,Myr (if pre-MS) or $90-350$\,Myr (if post-MS). Although the post-MS age, which would be too high, is more likely, the pre-MS cannot be excluded. Furthermore, with $2\,\sigma$ error bars it could also be young enough if it is post-MS. Therefore, we still consider HD\,76060 as a promising candidate.

As the association to PSR J0855$-$4644 is unlikely, we made a search for DR2 candidates which trace back to a position around the GC according to Eqn.~\ref{SNRerror} and found twelve objects with $G<17$\,mag which are considered good runaway candidates together with HD\,76060 and are suggested for follow-up observations.

\subsection{Constraints on runaway star production}

Here, we want to give an estimate how the findings of this work fit model populations of runaway stars in the literature (Bo17, Rz19). We investigated twelve SNRs and can exclude runaway stars with $G<17$\,mag in three of them. For SNR S147, we can confirm HD\,37424 with \textit{Gaia} DR2 data. Based on this association, we find no further runaway stars within S147. In each of the remaining eight SNRs, we find one or more candidates, but none of those can be confirmed yet.

So the minimum number of BES runaway stars within the twelve investigated SNRs is one, while the maximum number is nine, corresponding to a fraction of ejected runaway stars per SNR of $8-75$\,\%. The upper limit could be even higher if we also count the cases where more than one runaway star could have been ejected. In three of the SNRs (Cygnus Loop, Vela Jr., Lupus Loop), two of the remaining candidates could have a common origin. Although many more error ellipses of the runaway star candidates are overlapping, it was considered that for a certain age of the SNR the error regimes would shrink correspondingly.

This compares to theoretical values for runaway stars ejected from core-collapse SNe between 32.5\,\% (Bo17) and 68\,\% (Rz19). The latter value comes from $78_{-22}^{+9}$\,\% of binary systems that do not merge before the first SN, multiplied by $86_{-22}^{+10}$\,\% that get disrupted during the SN, which gives $67_{-26}^{+17}$\,\% of high-mass binary systems that eject a runaway star. Furthermore, Bo17 state that $\sim$1.22 core-collapse SNe happen in each high-mass binary system, so we get $55_{-21}^{+13}$\,\% of core-collapse SNe that eject a runaway star.

Both values are consistent with our findings. We need follow-up observations of our best candidates to search for SN debris as well as more precise estimates for the ages, distances and explosion sites of the SNRs to further constrain the numbers of ejected runaway stars in SNRs. From the theoretical values, we expect to verify $3-7$ of our runaway star candidates within the eleven SNRs besides S147 in the future.

\section{Conclusions}\label{sec6}

Our search in the \textit{Gaia} DR2 catalogue for runaway stars in twelve Galactic SNRs yielded no further certain runaway stars besides HD\,37424, but 73 new promising candidates, which are listed together with HD\,37424 in Table~\ref{Runawaylist}\hspace{-1.7mm}. Among these, five stars in the Monoceros Loop and one in Vela Jr. were observed with UVES/VLT and four stars in HB9 with Subaru/HDS.

In total, we observed 39 stars, where 29 were ruled out, e.g. after revisiting their motion with \textit{Gaia} DR2 data or they turned out to be giants. We found six dwarf stars with lithium; two in Monoceros, two in Vela and two in Vela Jr. TYC159-343-1, located in the Monoceros Loop in projection, might be a young star. We obtained $90-625$\,Myr from Fig~\ref{Mamajek}\hspace{-1.7mm} but its Li signal is hard to interpret due to the early spectral type (F$3\pm1$) \citep{Dantona1984}, and its low metallicity ([M/H]\,$=-0.57\pm0.08$) \citep{Lambert2004}. Even $2.1-3.8$\,Myr, the pre-MS solution from Fig.~\ref{HRD}\hspace{-1.7mm}, is possible. However, using DR2 data and an updated distance of the Monoceros Loop, TYC159-343-1 had to be ruled out because it is too close. The same applies to TYC159-251-1, which was discovered by us to be a double-lined spectroscopic binary. TYC8150-3105-1 in Vela shows the strongest Li absorption among all observed stars, so it was found to be younger than $\sim$50\,Myr. The Li-rich stars in Vela (TYC8150-3105-1 and TYC8150-2802-1) were ruled out because they are not consistent with the motion of the PSR. They could be interlopers near the volume of the SNR while TYC8150-3105-1 could also be a member of the same OB association from which the SNR formed. The Li-rich stars in Vela Jr. were ruled out because they are too old, which can be seen from their relatively small Li abundances.

The B-type star HD\,261393 is the most likely candidate in the Monoceros Loop. In Vela, we point out two stars which are consistent with the motion of the Vela PSR. The star marked in blue in Fig.~\ref{Velacands}\hspace{-1.7mm} is suggested for observations, while \textit{Star A} (FB19) is too faint to be investigated spectroscopically.

In total, we found 74 good runaway star candidates with $G<17$\,mag, mostly with spectral type K. We suggest to take high-resolution spectra to search for the Li 6708\,\AA{} line. Among the twelve investigated SNRs, nine SNe could have ejected a runaway star, while only HD\,37424 in S147 is confirmed. Three of these SNe could possibly have ejected two runaway stars.

Note that we might miss some stars with $G>17$\,mag which would mainly be spectral type M. Due to missing observations of such late-type stars around high-mass primaries, we do not know how many M-type runaway stars we could expect. Furthermore, we might miss some \textit{Gaia} DR2 candidates due to our strict selection limit from the allowed angular distance to the geometric centre (GC) at the time of the SN, which was necessary to create feasible observing projects.

In order to finally proof a BES origin, we need to observe the best candidates with very high resolution and $S/N$ to be able to detect SN debris, e.g. heavy- and $\alpha$-elements, in the stellar atmospheres. We also emphasise that it is important to precisely know the explosion site. The case of HD\,37424 in SNR S147 is rather exceptional, because both the pulsar and the runaway star trace back to the Green centre and meet it within just a few arcminutes. We did not find other cases where a pair of a pulsar and a runaway star candidate was located so close to the GC. In general, the explosion site will not coincide with the GC, due to the motion of the local standard of rest and due to asymmetries of the SNR expansion (e.g. \citealt{Meyer2015,Meyer2020}). A careful analysis of the time-dependent SNR morphology is needed to locate the explosion site.

\section*{Acknowledgements}

We want to thank our present and former colleagues for many hints and fruitful discussions, in particular Baha Din\c{c}el, Anna Pannicke and Christian Adam.
We acknowledge Sergi Blanco-Cuaresma for many useful hints for the usage of \textit{iSpec} and the analysis of stellar spectra in general; Eric Mamajek for providing the data and fits for $T_{\text{eff}}$ versus $EW_{\text{Li}}$ for clusters of different ages; and Aaron Steinhauer for providing the data that relate $\log(N_{\text{Li}})$, $v_{\text{rot}} \sin(i)$ and age. This work made use of ESO VLT data from run 0100.D-0314 and of NAOJ Subaru data from run S18B0195S. Therefore, we would like to thank the ESO and NAOJ staff, in particular the local support astronomers as well as John Pritchard (ESO), Chie Yoshida (NAOJ) and Akito Tajitsu (NAOJ), for the observations and many useful hints and comments about the data reduction. The NAOJ Subaru telescope is located on the Mauna Kea and we would like to acknowledge the important cultural and spiritual role, which the mountain has for the indigenous Hawaiian community.
This work made extensive use of data from the ESA-mission \textit{Gaia}\footnote{\href{https://www.cosmos.esa.int/gaia}{https://www.cosmos.esa.int/gaia}}, which were processed and provided by the \textit{Gaia Data Processing and Analysis Consortium} (DPAC). We further used \textit{VizieR, Simbad} and \textit{Aladin}, provided by the \textit{Centre de Donn\'{e}e astronomiques de Strasbourg} (CDS), and the \textit{Australia Telescope National Facility} (ATNF) Pulsar Catalogue \citep{Manchester2005}.

Funding was provided by the \textbf{Deutsche Forschungsgemeinschaft}, project \textbf{NE 515/57-1}.

\nocite{*}
\bibliography{Lux}

\appendix

\section{Overview of the runaway candidates\label{app2}}

Table~\ref{Obslist}\hspace{-1.7mm} shows the individual stars that were observed with UVES/VLT in P100 and with HDS/Subaru in S18B, respectively. All observations were executed in service mode. For the HDS spectra, the $S/N$ was measured in continuum regions between 6620\,\AA{} and 6760\,\AA{} in the combined spectra with \textit{IRAF}, while for the UVES spectra we obtained the $S/N$ from the ESO Archive Science Portal\footnote{\href{http://archive.eso.org/scienceportal/home}{http://archive.eso.org/scienceportal/home}}. Julian date and barycentric Julian date were calculated with the online tools from AAVSO\footnote{\href{https://www.aavso.org/jd-calculator}{https://www.aavso.org/jd-calculator}} and \citet{Eastman2010}\footnote{\href{http://astroutils.astronomy.ohio-state.edu/time/utc2bjd.html}{http://astroutils.astronomy.ohio-state.edu/time/utc2bjd.html}}, respectively.

Table~\ref{Runawaylist}\hspace{-1.7mm} gives an overview of the 74 possible runaway stars that were identified during this analysis. The spectral types SpT were estimated from the \textit{Gaia} DR2 $T_{\text{eff}}$ if available, with the following exceptions: HD\,37424 was observed, described and confirmed as a runaway star by \citet{Dincel2015}. The stars in SNR G205.5+00.5 as well as HD76060 in G266.2$-$01.2 were observed by us with UVES and their spectral types were adopted from the \textit{Skiff} catalogue (\citealt{Skiff2009}; \citealt{Houk1978}; Vo85). The stars in G160.9+02.6 (except the faintest one) were observed by us with HDS and their spectral types were adopted from the $T_{\text{eff}}$ given in \citet{Bai2019}.

\begin{center}
\begin{table*}%
\caption{List of observed stars. The columns give target name, instrument (U for UVES/VLT; H for HDS/Subaru), equatorial coordinates (J2000) calculated with \textit{VizieR} from \textit{Gaia} DR2 data, the SNR where the target is located, \textit{Gaia} DR2 \textit{G} magnitude, barycentric Julian date BJD, total on source integration time $t_{\text{exp}}$ and the $S/N$ reached in the fully reduced and averaged spectra.}
\label{Obslist}
\centering
\begin{tabular*}{490pt}{ccccccccc}
\hline
Name & Instr. & RA & DEC & SNR & $G$ & BJD$-2450000$ & $t_{\text{exp}}$ & $S/N$ \\
 & & [h:m:s] & [d:m:s] & & [mag] & [d] & [s] & \\
\hline
TYC1869-1435-1 & U & 05:38:15.15 & $+$27:49:18.48 & G180.0$-$01.7 &   12.20 & 8056.84606 & 900  & 112\\
TYC1869-1596-1 & U & 05:38:23.71 & $+$27:41:33.07 & G180.0$-$01.7 &   12.16 & 8060.80811 & 800  & 106\\
TYC1869-1670-1 & U & 05:39:03.05 & $+$27:32:35.26 & G180.0$-$01.7 &   12.57 & 8058.82311 & 900  & 87\\
TYC159-2540-1  & U & 06:37:52.95 & $+$06:39:53.48 & G205.5$+$00.5 &     12.73 & 8060.84497 & 300  & 48\\
HD\,260990     & U & 06:37:57.18 & $+$06:28:47.47 & G205.5$+$00.5 &     11.63 & 8075.79833 & 300  & 76\\
HD\,261117     & U & 06:38:19.74 & $+$06:23:40.70 & G205.5$+$00.5 &     9.74 &  8036.85181 & 60   & 81\\
TYC159-241-1   & U & 06:38:25.23 & $+$06:11:31.56 & G205.5$+$00.5 &     10.47 & 8076.79745 & 240  & 104\\
TYC159-2006-1  & U & 06:38:32.24 & $+$06:39:02.91 & G205.5$+$00.5 &     9.28 &  8076.80388 & 60   & 95\\
TYC159-2408-1  & U & 06:38:32.28 & $+$06:30:22.52 & G205.5$+$00.5 &     9.94 &  8076.80850 & 140  & 103\\
TYC159-2771-1  & U & 06:38:50.77 & $+$06:35:26.03 & G205.5$+$00.5 &     12.01 & 8076.81354 & 480  & 69\\
TYC159-1896-1  & U & 06:38:51.46 & $+$06:39:15.88 & G205.5$+$00.5 &     11.48 & 8076.82462 & 360  & 85\\
TYC159-2082-1  & U & 06:38:59.13 & $+$06:32:37.00 & G205.5$+$00.5 &     11.08 & 8076.83232 & 300  & 94\\
HD\,261359     & U & 06:39:07.47 & $+$06:27:37.99 & G205.5$+$00.5 &     11.79 & 8067.76600 & 720  & 114\\
HD\,261393     & U & 06:39:13.09 & $+$06:37:53.98 & G205.5$+$00.5 &     10.05 & 8060.81954 & 120  & 110\\
TYC159-2337-1  & U & 06:39:20.12 & $+$06:21:00.47 & G205.5$+$00.5 &     11.80 & 8107.61880 & 800  & 109\\
TYC159-2671-1  & U & 06:39:24.25 & $+$06:33:08.53 & G205.5$+$00.6 &     12.25 & 8076.84023 & 480  & 64\\
TYC159-892-1   & U & 06:39:24.38 & $+$06:37:25.98 & G205.5$+$00.5 &     9.44 &  8107.65051 & 100  & 109\\
HD\,261527     & U & 06:39:33.69 & $+$06:06:58.87 & G205.5$+$00.5 &     8.41 &  8107.65524 & 20   & 83\\
TYC159-2962-1  & U & 06:39:36.43 & $+$06:16:36.10 & G205.5$+$00.5 &     9.30 &  8107.65808 & 100  & 106\\
HD\,261589     & U & 06:39:44.89 & $+$06:27:35.57 & G205.5$+$00.5 &     11.06 & 8107.64301 & 360  & 99\\
TYC159-251-1   & U & 06:39:47.09 & $+$06:11:16.54 & G205.5$+$00.5 &     11.51 & 8107.66221 & 360  & 83\\
TYC159-2564-1  & U & 06:40:00.11 & $+$06:17:36.17 & G205.5$+$00.5 &     11.97 & 8107.63233 & 600  & 80\\
HD\,261715     & U & 06:40:13.41 & $+$06:16:45.02 & G205.5$+$00.5 &     11.25 & 8107.66940 & 480  & 98\\
TYC159-343-1   & U & 06:40:15.57 & $+$06:11:36.61 & G205.5$+$00.5 &     10.87 & 8107.67796 & 180  & 75\\
TYC7669-1336-1 & U & 08:22:05.66 & $-$42:58:08.41 & G260.4$-$03.4 & 12.27 & 8078.79809 & 1000 & 100 \\
TYC7669-1414-1 & U & 08:22:10.86 & $-$43:04:17.62 & G260.4$-$03.4 & 11.97 & 8078.81243 & 800  & 94\\
TYC8150-2802-1 & U & 08:34:02.97 & $-$45:32:59.68 & G263.9$-$03.3 & 12.03 & 8078.82413 & 1000 & 107 \\
TYC8150-3105-1 & U & 08:34:35.19 & $-$45:11:37.87 & G263.9$-$03.3 & 11.94 & 8078.83783 & 600  & 90\\
TYC8152-120-1  & U & 08:52:02.14 & $-$46:23:22.99 & G266.2$-$01.2 & 11.58 & 8107.68437 & 600  & 80\\
HD\,76060      & U & 08:52:02.45 & $-$46:17:19.84 & G266.2$-$01.2 & 7.85 &  8035.85245 & 20   & 116\\
TYC8152-1456-1 & U & 08:52:05.71 & $-$46:13:37.61 & G266.2$-$01.2 & 12.47 & 8077.80900 & 720  & 72\\
TYC8152-104-1  & U & 08:52:08.05 & $-$46:12:09.38 & G266.2$-$01.2 & 11.53 & 8077.82179 & 480  & 91\\
TYC8152-550-1  & U & 08:52:14.68 & $-$46:14:50.00 & G266.2$-$01.2 & 11.92 & 8077.83090 & 480  & 80\\
TYC3344-235-1  & H & 05:00:31.29 & $+$46:33:27.25 & G160.9$+$02.6 &     11.07 & 8418.99274 & 600   & 145 \\
TYC3344-679-1  & H & 05:00:53.98 & $+$46:30:00.61 & G160.9$+$02.6 &     11.40 & 8419.00315 & 600   & 151 \\
TYC3344-771-1  & H & 05:01:10.74 & $+$46:42:11.04 & G160.9$+$02.6 &     10.60 & 8419.01078 & 360   & 185 \\
TYC3344-683-1  & H & 05:01:10.87 & $+$46:27:38.11 & G160.9$+$02.6 &     12.33 & 8419.02190 & 1200  & 142 \\
TYC3344-124-1  & H & 05:01:11.84 & $+$46:47:57.30 & G160.9$+$02.6 &     10.45 & 8419.03231 & 240   & 136 \\
TYC3344-553-1  & H & 05:01:16.00 & $+$46:33:21.68 & G160.9$+$02.6 &     10.87 & 8419.03787 & 360   & 139 \\
\hline
\vspace{15mm}
\end{tabular*}
\end{table*}
\end{center}

\begin{center}
\begin{table*}
\caption{List of identified runaway star candidates. Equatorial coordinates (J2000) were calculated with \textit{VizieR} from \textit{Gaia} DR2 data. The fifth column shows the angular distance $\rho$ to the geometric centre at the time of the SN, except for the candidate in Vela (G263.9$-$03.3), where the distance to the location of the Vela pulsar at the time of the SN is given. The sixth column shows the $G$ magnitudes, taken from \textit{Gaia} DR2, and the last column shows the spectral types SpT.}
\label{Runawaylist}
\begin{tabular*}{470pt}{ccccccc}
\hline
Gaia DR2 & RA [h:m:s] & DEC [d:m:s] & SNR & $\rho$ [arcmin] & $G$ [mag] & SpT \\
\hline
1859462217726265728     & 20:50:22.21 & $+$30:30:17.37   & G074.0$-$08.5 &  $2.3\pm4.7$ & 16.23 & K2--K3.5 \\
1859461771049581952     & 20:50:47.80 & $+$30:30:43.62   & G074.0$-$08.5 &  $3.4\pm5.1$ & 16.88 & K5--K6 \\
1859469532044083968     & 20:50:54.62 & $+$30:39:06.92   & G074.0$-$08.5 &  $3.5\pm3.7$ & 15.20 & K2--K3 \\
1859469364551815040     & 20:50:54.66 & $+$30:37:29.63   & G074.0$-$08.5 &  $3.4\pm3.8$ & 15.65 & K2.5--K3.5 \\
1859469467631489920     & 20:50:55.57 & $+$30:38:13.63   & G074.0$-$08.5 &  $1.1\pm3.5$ & 15.47 & K2--K3.5 \\
1858709945619185664     & 20:50:56.65 & $+$30:26:27.61   & G074.0$-$08.5 &  $4.9\pm5.9$ & 15.06 & G9--K2 \\
1859471357406309120     & 20:50:57.42 & $+$30:42:38.52   & G074.0$-$08.5 &  $2.6\pm4.3$ & 14.44 & --         \\
1859471254338021120     & 20:50:58.70 & $+$30:40:51.90   & G074.0$-$08.5 &  $0.6\pm3.4$ & 16.67 & K1.5--K3.5 \\
1859471288697195136     & 20:50:59.09 & $+$30:41:00.61   & G074.0$-$08.5 &  $1.1\pm4.3$ & 14.22 & G9 \\
1859471082528259840     & 20:51:00.80 & $+$30:40:47.78   & G074.0$-$08.5 &  $1.7\pm4.2$ & 15.94 & G9--K3.5 \\
1859470975153131904     & 20:51:01.85 & $+$30:39:17.77   & G074.0$-$08.5 &  $2.0\pm4.2$ & 14.04 & G0--G3 \\
1859474823444598144     & 20:51:02.39 & $+$30:44:36.83   & G074.0$-$08.5 &  $3.7\pm4.1$ & 15.67 & K2--K3.5 \\
1859471185617972864     & 20:51:03.14 & $+$30:41:42.71   & G074.0$-$08.5 &  $3.5\pm4.3$ & 16.27 & K2.5--K3.5 \\
1859471872812767104     & 20:51:04.07 & $+$30:43:47.86   & G074.0$-$08.5 &  $1.9\pm4.1$ & 15.72 & K1.5--K3 \\
1859471219977693056     & 20:51:09.28 & $+$30:42:10.58   & G074.0$-$08.5 &  $1.7\pm4.2$ & 16.88 & K5--M0 \\
1858715683695452800     & 20:51:10.06 & $+$30:26:53.64   & G074.0$-$08.5 &  $1.7\pm6.8$ & 13.01 & G2--K0 \\
1858719291463059840     & 20:51:17.37 & $+$30:36:47.21   & G074.0$-$08.5 &  $1.7\pm3.5$ & 16.95 & F2--K6 \\
1858719085309596672     & 20:51:30.74 & $+$30:36:47.96   & G074.0$-$08.5 &  $4.0\pm3.6$ & 13.84 & G9--K3.5 \\
1859572714338261504     & 20:51:35.72 & $+$30:58:15.42   & G074.0$-$08.5 &  $6.8\pm6.5$ & 15.02 & K3--K6 \\
206966712992657792$^1$  & 05:00:31.29 & $+$46:33:27.25   & G160.9$+$02.6   & \hspace{1mm} $8  \pm14$    & 11.07 & A5--F0 \\
206918781157635072$^2$  & 05:00:53.98 & $+$46:30:00.61   & G160.9$+$02.6   &  $10 \pm17$   & 11.40 & A6--F0 \\
206994059046621824      & 05:00:55.07 & $+$46:40:37.90   & G160.9$+$02.6   &  $1.7\pm3.0$ & 16.27 & K5.5--K7 \\
206938263129288704$^3$  & 05:01:10.87 & $+$46:27:38.11   & G160.9$+$02.6   &  $12 \pm17$   & 12.33 & A7--F0 \\
206942519437580416$^4$  & 05:01:16.00 & $+$46:33:21.68   & G160.9$+$02.6   & \hspace{1mm} $7  \pm16$    & 10.87 & A7--A9 \\
3442490264261174528$^5$ & 05:39:44.40 & $+$27:46:51.19   & G180.0$-$01.7 &  $4.8\pm7.0$ & 8.91 & B0.5 \\
3133462316432771200$^6$ & 06:38:50.77 & $+$06:35:26.03   & G205.5$+$00.5   &  $12 \pm26$ & 12.01 & B8--A0 \\
3133412018076162560$^7$ & 06:39:07.47 & $+$06:27:37.99   & G205.5$+$00.5   & \hspace{1mm} $3  \pm22$ & 11.79 & B8--A0 \\
3133486093371718656$^8$ & 06:39:13.09 & $+$06:37:53.98   & G205.5$+$00.5   &  $10 \pm25$ & 10.05 & B4--B6 \\
3133407104633600512$^9$ & 06:39:20.12 & $+$06:21:00.47   & G205.5$+$00.5   &  $11 \pm22$ & 11.80 & B8--A0 \\
3133414491977266816$^{10}$ & 06:39:24.25 & $+$06:33:08.53  & G205.5$+$00.5 & \hspace{1mm} $7 \pm20$  & 12.25 & B7--B9 \\
5526325078219949440     & 08:22:00.32 & $-$42:59:34.28 & G260.4$-$03.4 &  $1.6\pm1.8$ & 13.58 & G2--G9     \\
5526328136236654208     & 08:22:00.50 & $-$42:57:26.84 & G260.4$-$03.4 &  $2.1\pm2.7$ & 13.17 & K1--K3   \\
5526325112579690496     & 08:22:05.08 & $-$42:58:43.34 & G260.4$-$03.4 &  $0.9\pm2.5$ & 16.86 & K7--M0     \\
5526324803342055168     & 08:22:05.62 & $-$43:01:23.19 & G260.4$-$03.4 &  $1.8\pm2.8$ & 16.92 & M1--M2     \\
5526324906421263744     & 08:22:06.33 & $-$42:59:23.89 & G260.4$-$03.4 &  $0.5\pm2.1$ & 15.47 & K3.5--K5.5 \\
5526325284378390784     & 08:22:10.14 & $-$42:59:29.27 & G260.4$-$03.4 &  $0.3\pm2.5$ & 14.87 & K5.5--K7   \\
5526324833397363840     & 08:22:10.47 & $-$43:00:49.73 & G260.4$-$03.4 &  $1.2\pm2.8$ & 15.99 & K5.5--M0   \\
5526325250018655616     & 08:22:15.57 & $-$42:58:31.52 & G260.4$-$03.4 &  $1.7\pm2.4$ & 13.56 & G9--K3     \\
5526325353097874176     & 08:22:17.60 & $-$42:58:40.09 & G260.4$-$03.4 &  $1.9\pm2.2$ & 13.00 & F2--F9     \\
5521967782363042432     & 08:35:48.51 & $-$45:18:46.12 & G263.9$-$03.3 &  $5.1\pm5.4$ & 15.85 & K3.5--M1.5 \\
5329655463417782528     & 08:51:46.03 & $-$46:21:01.11 & G266.2$-$01.2 &  $2.1\pm2.9$ & 16.38 & K4.5--K7   \\
5329655257259343232     & 08:51:47.10 & $-$46:21:17.64 & G266.2$-$01.2 &  $2.6\pm3.8$ & 16.87 & K6--K9     \\
\hline
\end{tabular*}
\begin{tablenotes}
\item $^1$TYC3344-235-1; $^2$TYC3344-679-1; $^3$TYC3344-683-1; $^4$TYC3344-553-1; $^5$HD\,37424; $^6$TYC159-2771-1; $^7$HD\,261359; $^8$HD\,261393; $^9$TYC159-2337-1; $^{10}$TYC159-2671-1
\end{tablenotes}
\end{table*}

\setcounter{table}{1}

\begin{table*}
\caption{Continued}
\begin{tabular*}{470pt}{ccccccc}
\hline
Gaia DR2 & RA [h:m:s] & DEC [d:m:s] & SNR & $\rho$ [arcmin] & $G$ [mag] & SpT \\
\hline
5329655845686016256     & 08:51:50.00 & $-$46:20:20.25 & G266.2$-$01.2 &    $1.7\pm3.4$ & 15.13 & K4--K5.5   \\
5329655983110009856     & 08:51:53.68 & $-$46:18:19.05 & G266.2$-$01.2 &    $1.7\pm3.9$ & 16.57 & K6--K9     \\
5329655776966532864     & 08:51:54.73 & $-$46:19:37.86 & G266.2$-$01.2 &    $0.7\pm2.8$ & 15.24 & K2.5--K3.5 \\
5329657563672287872     & 08:51:55.40 & $-$46:16:42.01 & G266.2$-$01.2 &    $2.8\pm4.3$ & 16.26 & K5--K8     \\
5329656017469763456     & 08:51:59.13 & $-$46:18:09.17 & G266.2$-$01.2 &    $1.0\pm4.1$ & 15.30 & K2--K4     \\
5329655811325638272     & 08:51:59.71 & $-$46:19:01.93 & G266.2$-$01.2 &    $0.7\pm4.1$ & 14.91 & G8--K2     \\
5329657288780088704     & 08:52:00.36 & $-$46:17:57.93 & G266.2$-$01.2 &    $1.8\pm4.3$ & 16.62 & K2.5--K6.5 \\
5329654849253600896     & 08:52:00.56 & $-$46:21:49.15 & G266.2$-$01.2 &    $2.9\pm4.3$ & 14.97 & K1.5--K5   \\
5329654849241704320     & 08:52:00.64 & $-$46:21:48.85 & G266.2$-$01.2 &    $2.5\pm4.3$ & 16.82 & --         \\
5329657490638374272$^{11}$ & 08:52:02.45 & $-$46:17:19.84 & G266.2$-$01.2 & $2.2\pm4.1$ & 7.85 & B8\,IV,V \\
5329654883601436032     & 08:52:05.09 & $-$46:22:19.13 & G266.2$-$01.2 &    $2.5\pm4.0$ & 14.91 & K0.5--K2.5 \\
6005336183671977344     & 15:09:15.49 & $-$40:00:55.10 & G330.0$+$15.0   &  $1.1\pm4.4$ & 16.31 & K7--M0 \\
6005322405417134848     & 15:09:22.17 & $-$40:17:50.46 & G330.0$+$15.0   &  $4.3\pm6.9$ & 15.95 & K3.5--M1 \\
6005309932830763648     & 15:09:27.22 & $-$40:24:26.21 & G330.0$+$15.0   &  $1.7\pm4.1$ & 15.70 & K9--M2.5 \\
5972266305587974656 & 17:13:45.20 & $-$39:46:11.77 & G347.3$-$00.5 &        $1.3\pm2.5$ & 14.71 & K1.5--K3.5 \\
5972266443017191680 & 17:13:47.23 & $-$39:44:36.86 & G347.3$-$00.5 &        $0.7\pm2.4$ & 15.97 & K5.5--M0.5  \\
5972266683540067328 & 17:13:47.48 & $-$39:43:14.68 & G347.3$-$00.5 &        $1.9\pm2.8$ & 16.84 & K6--M0.5   \\
5972266443017192960 & 17:13:47.51 & $-$39:44:29.27 & G347.3$-$00.5 &        $0.7\pm2.6$ & 16.65 & K6--M1.5   \\
5972266447337050496 & 17:13:49.35 & $-$39:44:01.38 & G347.3$-$00.5 &        $1.1\pm2.9$ & 16.98 & K6--M1     \\
5972266447374127872 & 17:13:49.64 & $-$39:44:03.04 & G347.3$-$00.5 &        $1.0\pm2.9$ & 15.01 & K4--K6      \\
5972266378654585856 & 17:13:50.45 & $-$39:45:35.03 & G347.3$-$00.5 &        $0.5\pm2.8$ & 16.59 & K6.5--K9   \\
5972266477376942848 & 17:13:51.06 & $-$39:43:28.12 & G347.3$-$00.5 &        $1.8\pm2.9$ & 14.81 & K1--K4.5   \\
5972219340115696512 & 17:13:51.18 & $-$39:46:39.54 & G347.3$-$00.5 &        $1.7\pm2.9$ & 16.54 & --         \\
5972266378617336832 & 17:13:51.38 & $-$39:45:25.30 & G347.3$-$00.5 &        $0.5\pm2.6$ & 16.40 & K3.5--K6.5 \\
5972266481696795008 & 17:13:52.37 & $-$39:43:28.21 & G347.3$-$00.5 &        $1.6\pm2.8$ & 16.92 & K3.5--K9   \\
5972266408657455872 & 17:13:53.26 & $-$39:44:22.17 & G347.3$-$00.5 &        $0.9\pm2.5$ & 16.88 & K5--M1     \\
5972266413014335616 & 17:13:53.46 & $-$39:44:41.28 & G347.3$-$00.5 &        $0.7\pm2.1$ & 16.59 & K7--M2.5   \\
5972219374522872576 & 17:13:54.47 & $-$39:46:45.27 & G347.3$-$00.5 &        $1.9\pm2.7$ & 15.04 & K4.5--K6    \\
5972219477602131840 & 17:13:55.22 & $-$39:45:21.84 & G347.3$-$00.5 &        $1.0\pm2.0$ & 13.47 & G8--K1.5   \\
5972266413014326528 & 17:13:55.45 & $-$39:44:36.52 & G347.3$-$00.5 &        $1.3\pm2.2$ & 15.62 & K5--K9     \\
5972267886141988864 & 17:13:56.10 & $-$39:44:03.01 & G347.3$-$00.5 &        $1.6\pm2.4$ & 13.94 & M3.5       \\
5972267890445877120 & 17:13:56.96 & $-$39:43:57.12 & G347.3$-$00.5 &        $1.8\pm2.4$ & 16.91 & K4.5--K9   \\
\hline
\end{tabular*}
\begin{tablenotes}
\item $^{11}$HD\,76060
\end{tablenotes}
\end{table*}
\end{center}

\section*{Author Biography}

\begin{biography}{}
{\textbf{Oliver Lux.}
The author obtained a Bachelor of Science in Physics and a Master of Science in Astrophysics at the Universities of Bochum and Bonn, respectively, gaining knowledge about neutron stars. In his current work as a PhD student at the University of Jena, he works on runaway stars and supernova remnants. Further research interests are variable stars and exoplanets.}
\end{biography}

\end{document}